\documentclass{aa}
\usepackage{graphicx}
\usepackage[varg]{txfonts}
\begin{document}

\newcommand{\be}{\begin{equation}}
\newcommand{\ee}{\end{equation}}

\title{Primordial nucleosynthesis with varying fundamental constants}
\subtitle{Degeneracies with cosmological parameters}

\author{C. J. A. P. Martins\inst{1,2}}
\institute{Centro de Astrof\'{\i}sica da Universidade do Porto, Rua das Estrelas, 4150-762 Porto, Portugal\\
\email{Carlos.Martins@astro.up.pt}
\and
Instituto de Astrof\'{\i}sica e Ci\^encias do Espa\c co, CAUP, Rua das Estrelas, 4150-762 Porto, Portugal}
\date{Submitted \today}

\abstract
{The success of primordial nucleosynthesis as a cornerstone of the hot Big Bang model has been limited by the long-standing lithium problem. Recent work presented a self-consistent perturbative analysis of the effects of variations in nature's fundamental constants on primordial nucleosynthesis for a broad class of grand unified theory models, showing that such models provide a possible solution to the lithium problem, provided the value of the fine-structure constant $\alpha$ at the nucleosynthesis epoch is larger than the current laboratory one by a few parts per million of relative variation. Here we extend the earlier analysis, focusing on how this preferred value of $\alpha$ is affected if relevant cosmological parameters are also allowed to vary--specifically focusing on the baryon-to-photon ratio, the number of neutrinos, and the neutron lifetime. We rephrase the lithium problem in terms of the values of these parameters that would be needed to solve it within this class of grand unified theories, thus obtaining values that would disagree with the results of other experiments by several standard deviations. Using these experimental results as priors in the analysis, we find that a larger value of $\alpha$ is still preferred, confirming our previous results. By excluding lithium from the analysis, we also obtain upper limits on possible variations of $\alpha$ at the primordial nucleosynthesis epoch. At the two-sigma level, these are $|\Delta\alpha/\alpha|<50$ ppm without nuclear physics, cosmology, or atomic clocks priors, or alternatively $|\Delta\alpha/\alpha|<5$ ppm if these priors are used. While the simplest solution to the lithium problem is likely to be found within observational astrophysics, our work shows that varying fundamental constants remain a viable alternative.}


\titlerunning{Primordial Nucleosynthesis with Varying Fundamental Constants}
\authorrunning{Martins}
\maketitle

%%%%%%%%%%%%%%%%%%%%%%%%%%%%%%%%%%%%%%%%%%%%%%%%%%%%%%%%%%%%%%%%%%%%%%%%%%
\section{Introduction}
\label{introd}

Big Bang nucleosynthesis (henceforth BBN) is one of the cornerstones of the standard particle cosmology paradigm. In its simplest form, and specifically assuming that the relevant nuclear physics is known, it has a single free parameter (the baryon-to-photon ratio), and can therefore tightly constrain physics beyond the standard model \citep{Steigman,Iocco,Pitrou}. Nevertheless, this is not an unqualified success story due to the well-known lithium problem \citep{Fields}. Indeed, the theoretically expected abundance of lithium-7 (given our present knowledge of astrophysics as well as nuclear and particle physics) exceeds the observed one by a factor of about 3.5; when allowing for the statistical uncertainties on both sides, this represents a detection of ignorance (or missing physics) at more than five standard deviations.

Attempts to solve the problem can be broadly classified into four categories--see \citet{Mathews} and the BBN section of the latest Particle Data Group (henceforth PDG) review by \citet{PDG} for more detailed, recent discussions. A mundane solution would involve systematics in astrophysical observations, although none have been identified so far that would account for the large difference that would be required for a solution. Analogously, one could envisage systematics on the nuclear physics side, specifically in the measurements of the required cross-sections; however, the steady improvements in experimental techniques have all but closed this possible loophole--a recent discussion can be found in \citet{Iliadis}. On the other hand, the problem could be due to unknown astrophysics, but again, no specific mechanism has been identified that can account for the discrepancy. Some degree of lithium depletion can occur in stars due to the mixing of the outer layers with the hotter interior \citep{Sbordone}, and some authors have suggested that a depletion by a factor as large as 1.8 may have occurred \citep{Ryan,Korn}. On the other hand, that scenario is difficult to reconcile with the existence of extremely iron-poor dwarf stars with lithium abundances, which are very close to the Spite plateau \citep{Aguado}.

Last but not least, the lithium problem could also point to new physics beyond the standard paradigm, including early or late-time decaying particles, for example \citep{Jedamzik,Cumberbatch,Pradler}. One bottleneck for possible solutions of this kind is that they often come at the cost of overproducing deuterium--indeed, for the observationally relevant ratios of the baryon-to-photon ratio, the two abundances are anticorrelated \citep{Anticorr1,Anticorr2}. This is also the case if one allows for a varying fine-structure constant, while assuming that no other gauge or Yukawa couplings are affected as has been done in several early studies \citep{Bergstrom,BBN0,Nollett,Ichikawa}.

In \citet{Clara} (henceforth CM20), we have recently proposed a solution belonging to the fourth category, relying on the spacetime variation of nature's fundamental couplings, which is unavoidable in most extensions of the standard model \citep{Damour,Uzan,ROPP}. Specifically, we developed a self-consistent perturbative methodology to study the impact on BBN of a broad but physically motivated class of extensions of the standard model--specifically grand unified theories (henceforth GUTs)--where all the gauge and Yukawa couplings are allowed to vary, bearing in mind that in each specific model within this class these variations are related to one another, but these relations themselves are model-dependent.

In the approach introduced in CM20, which builds upon earlier work by \citet{Muller} and \citet{Resonance1} and subsequent developments by \citet{Coc} and \citet{Stern1}, one can generically write the relative variations of other couplings as the product of some constant coefficients and the relative variation of the fine-structure constant, $\alpha=e^2/({\bar h}c)$, and with some reasonable simplifying assumptions \citep{Campbell}, only two such coefficients are needed: one pertaining to electroweak physics and the other to strong interactions. This has the further advantage of enabling an extensive phenomenological exploration of the possible models within this class.

Qualitatively one expects that the effects of the varying couplings on BBN will be larger for heavier nuclides. Therefore, it is plausible that a suitable amount of variation will reconcile the theoretical and observed lithium-7 abundances without significantly impacting those of the lighter nuclides--especially that of deuterium, for which very stringent observational constraints exist. In CM20 we have quantitatively confirmed this expectation, showing that there are models within this class where the lithium problem has indeed been solved, at least in the statistical sense that the observed and predicted abundances all agree to better than three standard deviations. All such models require a value of $\alpha$ at the BBN epoch that is larger than the present-day laboratory value. Astrophysical constraints on $\alpha$ are typically expressed in terms of a relative variation with respect to the laboratory value, $(\Delta\alpha/\alpha)(z)=(\alpha(z)-\alpha_0)/\alpha_0$. In this case, and although the preferred value is model-dependent, the required relative variation at the BBN epoch is at the level of a few parts per million (henceforth ppm). We note that these BBN results are two to three orders of magnitude more stringent than the ones obtained at redshift $z\sim1100$ from cosmic microwave background observations, which make the simplifying assumption that only $\alpha$ varies while the rest of the physics is standard \citep{Planckalpha,Hart} and that sub-ppm sensitivity has only recently been achieved for high-resolution spectroscopy measurements at low-density absorption clouds along the line-of-sight of bright low-redshift quasars \citep{Kotus}. In any case, such ppm-level variations at the BBN epoch would be consistent with all current constraints at lower redshifts. A recent review of these constraints and their implications for several cosmological and particle physics models can be found in \citet{ROPP}.

One limitation of the analysis in CM20 is that the perturbative analysis was applied to a baseline scenario where other relevant parameters (such as the baryon-to-photon ratio $\eta$ or the number of light neutrino species $N_\nu$) were kept fixed at their standard values rather than being allowed to vary themselves. While, in principle, this is perfectly consistent--there is no reason why varying couplings would affect these parameters by themselves-- in practice, it is possible that the best-fit values of $\alpha$ obtained in CM20 will change when this assumption is relaxed because the predicted abundances of the light elements will also depend on these additional parameters.\ In other words, there may be correlations between $\alpha$ and these parameters. The goal of the present work is to address this issue. In addition to the baryon-to-photon ratio or the number of light neutrino species, we also briefly consider the role of the neutron lifetime, which can be separately measured in local experiments.

The plan of the rest of this work is as follows. We start in Section \ref{guts} with a brief summary of the formalism and results of CM20, with the aim of making the present work self-contained. We then discuss the correlations between $\alpha$ and the neutron lifetime in Section \ref{neutron} and those between $\alpha$, the baryon-to-photon ratio, and the number of light neutrino species in Section \ref{cosmo}. Finally, we report our main conclusions in Section \ref{concl}.

%%%%%%%%%%%%%%%%%%%%%%%%%%%%%%%%%%%%%%%%%%%%%%%%%%%%%%%%%%%%%%%%%%%%%%%%%%%%%%
\section{Methodology and previous results}
\label{guts}

Here we provide a short but hopefully self-contained overview of the results in CM20, emphasising the points that are relevant for our present discussion. The reader is invited to refer to CM20 and further references therein for detailed derivations.

In order to describe models which allow for simultaneous variations of several fundamental couplings, one needs to relate the various changes to those of a particular dimensionless coupling, conveniently chosen to be $\alpha$. As in \citet{Coc}, we consider a broad class of grand unification models, where the weak scale is determined by dimensional transmutation, further assuming that the relative variation of all the Yukawa couplings is the same and that the variation of the couplings is driven by a dilaton-type scalar field \citep{Campbell}. These assumptions suffice to self-consistently relate the relative variations of all the parameters impacting BBN.

To give one specific example, since fundamental particle masses are the product of the Higgs vacuum expectation value $\nu$ and the corresponding Yukawa coupling $h$, the electron mass will vary as
\begin{equation}
\frac{\Delta m_e}{m_e}=\frac{1}{2}(1+S)~\frac{\Delta\alpha}{\alpha}\,,
\end{equation}
where $S$ is a dimensionless parameter related to electroweak physics, which is defined as
\begin{equation}
\frac{\Delta \nu}{\nu} = S \frac{\Delta h}{h}\,;
\end{equation}
we recall that here we assume that all the Yukawa couplings vary in the same way. In the case of the proton, which is a composite particle, one needs a second dimensionless parameter, $R$, related to quantum chromodynamics (QCD),
\begin{equation}
\frac{\Delta \Lambda_{QCD}}{\Lambda_{QCD}} = R \frac{\Delta \alpha}{\alpha}
\end{equation}
where $\Lambda_{QCD}$ is the QCD mass scale. This leads to a proton mass variation of
\begin{equation}
\frac{\Delta m_p}{m_p}=\big[0.8R+0.2(1+S) \big]~\frac{\Delta\alpha}{\alpha}\,.
\end{equation}
In this perturbative approach, analogous expressions for the relative variations of other relevant quantities can be obtained, including the neutron mass (and thus also the mass difference between neutrons and protons and the average nucleon mass), Newton's gravitational constant, the binding energies of relevant species (deuterium, tritium, helium-3, helium-4, lithium-7, and beryllium-7), and finally the neutron lifetime, $\tau_n$, which we write explicitly since it is relevant for what follows
\begin{equation}
\frac{\Delta \tau_n}{\tau_n}=[-0.2-2.0~S+3.8~R]\, \frac{\Delta\alpha}{\alpha}\,;
\end{equation}
the remaining expressions can be found in CM20.

The phenomenological parameters $R$ and $S$ can, in principle, be considered as free parameters to be experimentally or observationally constrained. Our current knowledge of particle physics and unification scenarios indicates that their absolute values can be anything from order unity to several hundreds, where $R$ is allowed to be positive or negative (though the former case is more likely), while $S$ is expected to be non-negative.

In some specific models, the two parameters can be calculated, and in what follows, we consider two specific cases, which have been shown in CM20 to provide solutions to the lithium problem. The first is a `typical' unification scenario, for which one has \citep{Coc,Langacker}
\begin{equation}
R\sim36\,,\quad S\sim160\,;
\end{equation}
we refer to this as the unification model. On the other hand, for the dilaton-type model discussed by \citet{Nakashima}, one instead has
\begin{equation}
R\sim109.4\,,\quad S\sim0\,;
\end{equation}
we refer to this as the dilaton model. As in CM20, we also consider the general case where the parameters $R$ and $S$ are allowed to vary and they are then marginalised. 

Generally, the sensitivity of the primordial abundances on the various relevant particle physics parameters can be described as \citep{Stern1}
\begin{equation}
\frac{\Delta Y_i}{Y_i}=\sum_j C_{ij}\frac{\Delta X_j}{X_j}\,,
\end{equation}
where $C_{ij}=\partial\ln{(Y_i)}/\partial\ln{(X_j)}$ are the sensitivity coefficients. The perturbation is always done with respect to the values in some baseline theoretical model, which is specified below. In our case, the sensitivities are expressed as a function of the unification parameters $R$ and $S$ and the relative variation of $\alpha$, in other words
\begin{equation}
\frac{\Delta Y_i}{Y_i}=(x_i+y_iS+z_iR)\frac{\Delta\alpha}{\alpha}\,;
\end{equation}
the sensitivity coefficients are listed in Table \ref{table1}.

%%%%%%%%%%%%%%%%%%%%%%%%%%%%%%%%%%%%%%%%%%%%%%%%%%%%%%%%%%%%%%%%%%%%%%%%%%%%%%
\begin{table}
\caption{Sensitivity coefficients of BBN nuclide abundances on the free parameters of our phenomenological parametrisation, defined in the main text. Reprinted from \citet{Clara}.}
\label{table1}
\centering
\begin{tabular}{| c | c c c c |}
\hline
$C_{ij}$ & D & ${}^3$He & ${}^4$He & ${}^7$Li \\
\hline
$x_i$ & 42.0 & 1.27 & -4.6 & -166.6 \\
$y_i$ & 39.2 & 0.72 & -5.0 & -151.6 \\
$z_i$ & 36.6 & -89.5 & 14.6 & -200.9 \\
\hline
\end{tabular}
\end{table}
%%%%%%%%%%%%%%%%%%%%%%%%%%%%%%%%%%%%%%%%%%%%%%%%%%%%%%%%%%%%%%%%%%%%%%%%%%%%%%

For our analysis, we need to specify a baseline standard model, yielding some theoretically predicted abundances, as well as the corresponding observationally measured values--with the two being compared using standard statistical likelihood analyses as well as theoretical and observational uncertainties being added in quadrature. For the baseline theoretical model, we use the values published in the recent review by \citet{Pitrou}. Specifically, the reference values for the neutron lifetime, the number of neutrinos, and the baryon-to-photon ratio are $\tau_n=879.5$, $N_\nu=3,$ and $\eta_{0}=6.1 $, respectively; these are further justified in the following sections. For the observed values, we rely on the recommended values in the latest PDG review \citep{PDG}. For convenience these values, which coincide with those used in CM20, are listed in Table \ref{table2}.

%%%%%%%%%%%%%%%%%%%%%%%%%%%%%%%%%%%%%%%%%%%%%%%%%%%%%%%%%%%%%%%%%%%%%%%%%%%%%%
\begin{table}
\caption{Theoretical and observed primordial abundances used in our analysis. The theoretical ones have been obtained in \citet{Pitrou}. The observed ones are the recommended values in the 2019 Particle Data Group BBN review, with a representative reference being \citet{Aver}, \citet{Cooke}, \citet{Bania}, and \citet{Sbordone}, respectively. These assumptions coincide with those in \citet{Clara}. We note that strictly speaking, the ${}^3He$ value is not cosmological; see the main text for further discussion.}
\label{table2}
\centering
\begin{tabular}{c c c}
\hline
Abundance & Theoretical & Observed \\
\hline
$Y_p$ & $0.24709\pm0.00017$ & $0.245\pm0.003$ \\
$(D/H)\times 10^5$ & $2.459\pm0.036$ & $2.545\pm0.025$ \\
$({}^3He/H)\times 10^5$ & $1.074\pm0.026$ & $1.1\pm0.2$ \\
$({}^7Li/H)\times 10^{10}$ & $5.624\pm0.245$ & $1.6\pm0.3$ \\
\hline
\end{tabular}
\end{table}
%%%%%%%%%%%%%%%%%%%%%%%%%%%%%%%%%%%%%%%%%%%%%%%%%%%%%%%%%%%%%%%%%%%%%%%%%%%%%%

The analysis of CM20 has shown that both the unification and dilaton models as defined above can provide a solution to the lithium problem for the best-fit values of $\Delta\alpha/\alpha$ listed in Table \ref{table3}. This table also shows the analogous results for the case where the parameters $R$ and $S$ are allowed to vary and subsequently marginalised, in this case with uniform priors in the range $R=[0,+500]$, $S=[0,+1000]$, together with an additional prior coming from local experiments with atomic clocks \citep{Clocks}
\begin{equation}\label{clopri}
(1+S)-2.7R=-5\pm15\,;
\end{equation}
in what follows, this is referred to as the clocks model. As pointed out in CM20, in the case of marginalised parameters, the results do depend on the choice of priors. Here we conservatively adopt this clocks case because among the choices of priors considered in CM20, it is the one that leads to a best-fit value of $\alpha$ which is closest to the null result; in this sense, it is therefore the most conservative choice out of the priors.

Finally, as in CM20, in what follows we present the results of analyses using three different combinations of the four abundances as follows.

Firstly, he baseline case uses the abundances of helium-4, deuterium, and lithium-7, which are the three available cosmological abundances. This should be seen as the analysis providing the reference values for the best-fit BBN values for $\alpha$.

Secondly, the extended case adds the helium-3 abundance to the former three; the reason for separately treating helium-3 is that its observed abundance is a local, rather than cosmological. In any case, as is pointed out in CM20--and as is subsequently confirmed in more detail in the present analysis--this has an extremely small impact on the derived constraints.

Thirdly, the null case uses the helium-4, deuterium, and helium-3 abundances, but it does not use lithium-7. The motivation for this case is that it provides a useful null test of the BBN sensitivity to the value of the fine-structure constant; in other words, it is expected that the standard value of $\alpha$ is recovered in this case to some degree of sensitivity that is useful to quantify and compare to other probes. It also provides an indication of the BBN constraints on $\alpha$ on the assumption that the lithium problem has an astrophysical solution.

For convenience, we use the terms, baseline, extended, and null, to denote each of the three combinations when presenting the results in figures and tables in the following sections.

%%%%%%%%%%%%%%%%%%%%%%%%%%%%%%%%%%%%%%%%%%%%%%%%%%%%%%%%%%%%%%%%%%%%%%%%%%%%%%
\begin{table}
\caption{Constraints on $\Delta\alpha/\alpha$ for the unification and dilaton models as well as for the general case with an atomic clocks prior (the clocks model), as reported in \citet{Clara}, for the various choices of primordial abundances used in the analysis. The listed values are in ppm, and they correspond to the best fit in each case and to the range of values within $\Delta\chi^2=4$ of it.}
\label{table3}
\centering
\begin{tabular}{c c c c}
\hline
Abundances & Unification & Dilaton & Clocks \\
\hline
Baseline & $12.5\pm2.9$ & $19.9\pm4.5$ & $2.2^{+15.6}_{-0.6}$ \\
Null & $4.6\pm3.8$ & $5.8\pm6.5$ & $1.0^{+7.1}_{-0.9}$ \\
Extended & $12.5\pm2.9$ & $19.5\pm4.5$ & $2.2^{+15.5}_{-0.6}$ \\
\hline
\end{tabular}
\end{table}
%%%%%%%%%%%%%%%%%%%%%%%%%%%%%%%%%%%%%%%%%%%%%%%%%%%%%%%%%%%%%%%%%%%%%%%%%%%%%%

%%%%%%%%%%%%%%%%%%%%%%%%%%%%%%%%%%%%%%%%%%%%%%%%%%%%%%%%%%%%%%%
\section{Neutron lifetime}
\label{neutron}

We start by revisiting the analysis in CM20  by separately treating the neutron lifetime, $\tau_n$. In this case the mathematical expression for the sensitivities becomes
\begin{equation}
\frac{\Delta Y_i}{Y_i}=(x_{n,i}+y_{n,i}S+z_{n,i}R)\frac{\Delta\alpha}{\alpha}+t_i\frac{\Delta\tau_n}{\tau_n}\,,
\end{equation}
where the additional $n$ subscript in the previously defined sensitivity coefficients $x$, $y$, and $z$ serves to distinguish them from the ones discussed in the previous section and is listed in Table \ref{table1}. Naturally we now have an additional set of sensitivity coefficients for the neutron lifetime, denoted $t_i$. It is straightforward to re-calculate the former sensitivity coefficients, while the ones for the neutron lifetime have been previously discussed in \citet{Pitrou} and also in \citet{Fields2}. All these sensitivity coefficients are listed in Table \ref{table4}.

%%%%%%%%%%%%%%%%%%%%%%%%%%%%%%%%%%%%%%%%%%%%%%%%%%%%%%%%%%%%%%%%%%%%%%%%%%%%%%
\begin{table}
\caption{Sensitivity coefficients of BBN nuclide abundances on the free parameters of our phenomenological parametrisation, defined in the main text, for the case when the neutron lifetime is treated as a separate parameter.}
\label{table4}
\centering
\begin{tabular}{| c | c c c c |}
\hline
$C_{ij}$ & D & ${}^3$He & ${}^4$He & ${}^7$Li \\
\hline
$x_{n,i}$ & 42.1 & 1.30 & -4.5 & -166.5 \\
$y_{n,i}$ & 40.1 & 2.00 & -3.5 & -150.7 \\
$z_{n,i}$ & 34.9 & -90.0 & 11.8 & -202.6 \\
$t_i$ & 0.442 & 0.141 & 0.732 & 0.438 \\
\hline
\end{tabular}
\end{table}
%%%%%%%%%%%%%%%%%%%%%%%%%%%%%%%%%%%%%%%%%%%%%%%%%%%%%%%%%%%%%%%%%%%%%%%%%%%%%%

The motivation for this analysis stems from the fact that neutron lifetimes can be experimentally measured. Indeed, two different experimental methods are commonly used, which are known as the bottle and beam methods. Broadly speaking, the first counts neutrons (in principle being sensitive to all decay channels), while the second counts protons and electrons produced by beta decay. A recent discussion can be found in \citet{Rajan}. A combination of the experimental results obtained with each of the methods gives
\be
\tau_{n,bottle}=879.6\pm0.6 \, s
\ee
\be
\tau_{n,beam}=888\pm2\, s\,,
\ee
which are formally discrepant at more than four standard deviations. In what follows, for consistency with the analysis in CM20, we use the value also adopted by \citet{Pitrou}
\be
\tau_n=879.5\pm0.8\, s\,,\label{neutronpri}
\ee
which also agrees with the value recommended by the latest PDG review \citep{PDG} of $\tau_n=879.4\pm0.6$ s.

\begin{figure*}
\centering
\includegraphics[width=8cm]{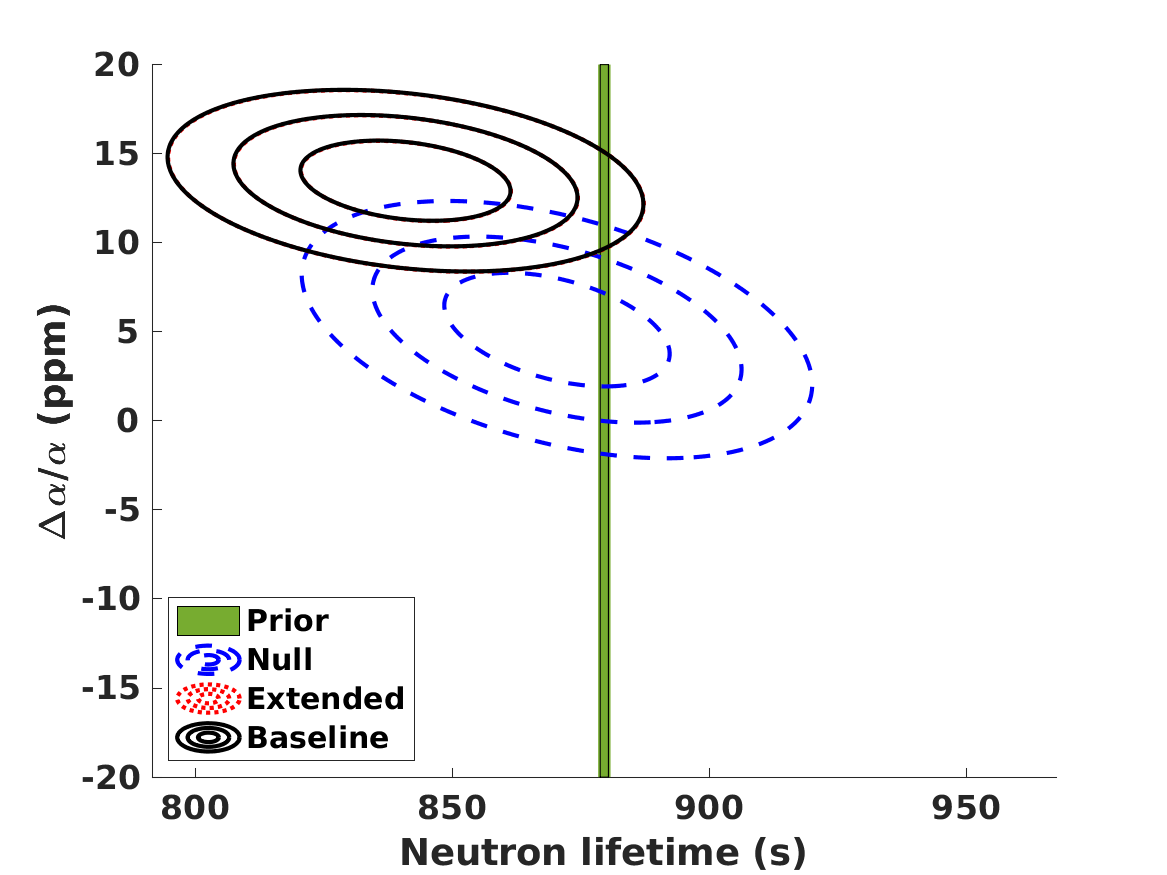}
\includegraphics[width=8cm]{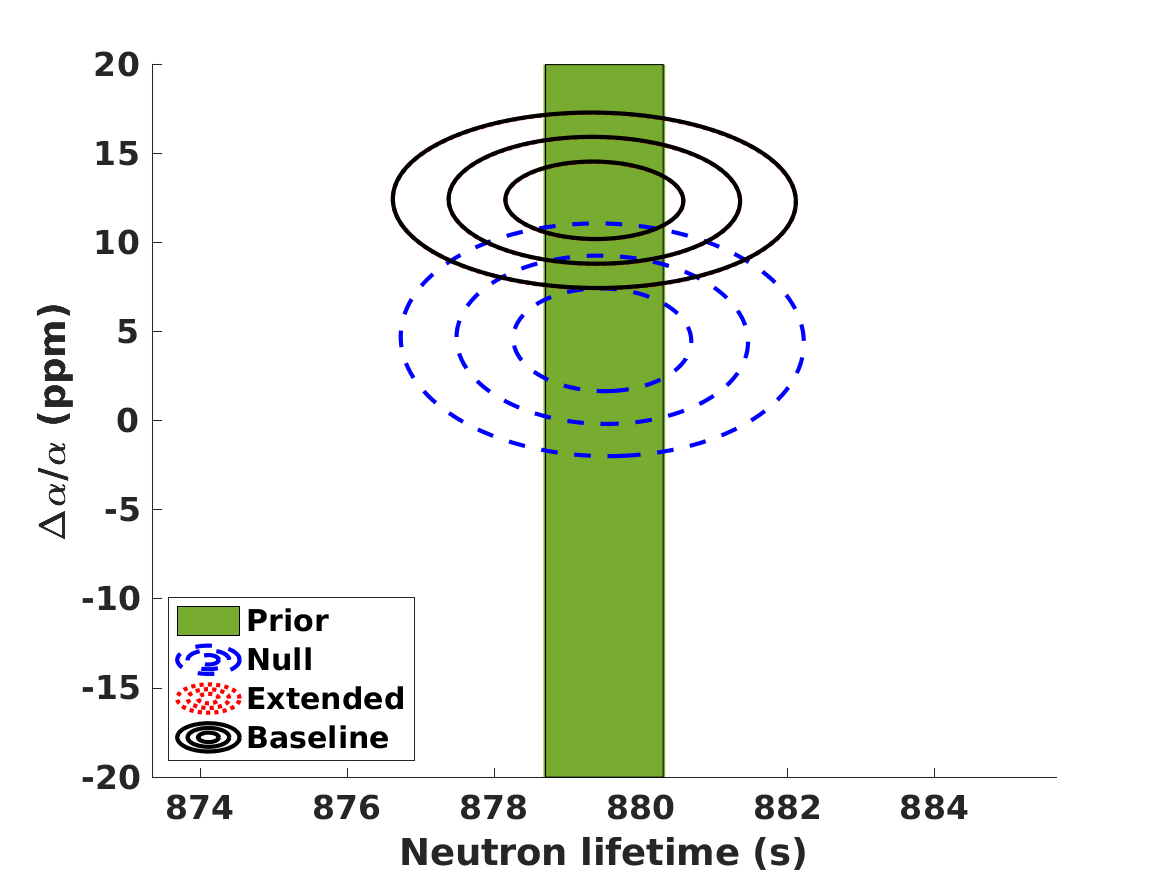}
\includegraphics[width=8cm]{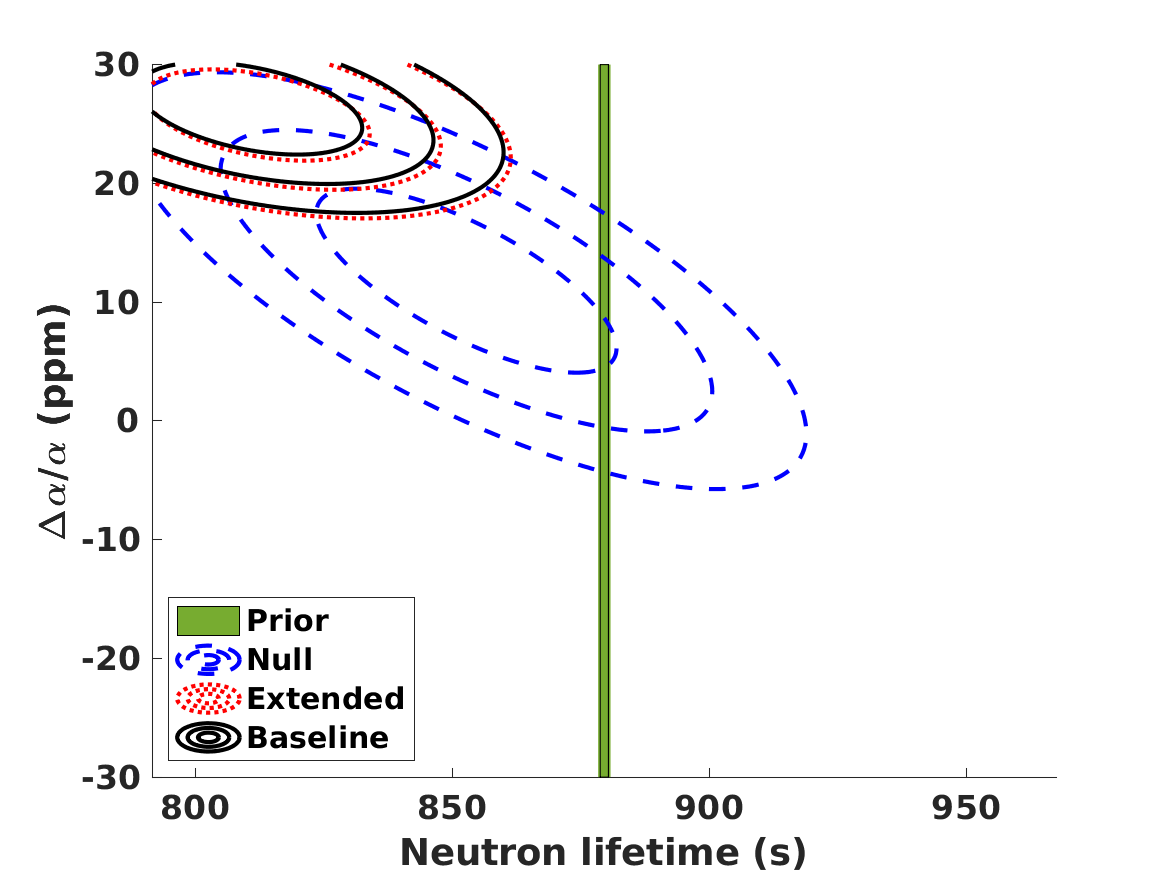}
\includegraphics[width=8cm]{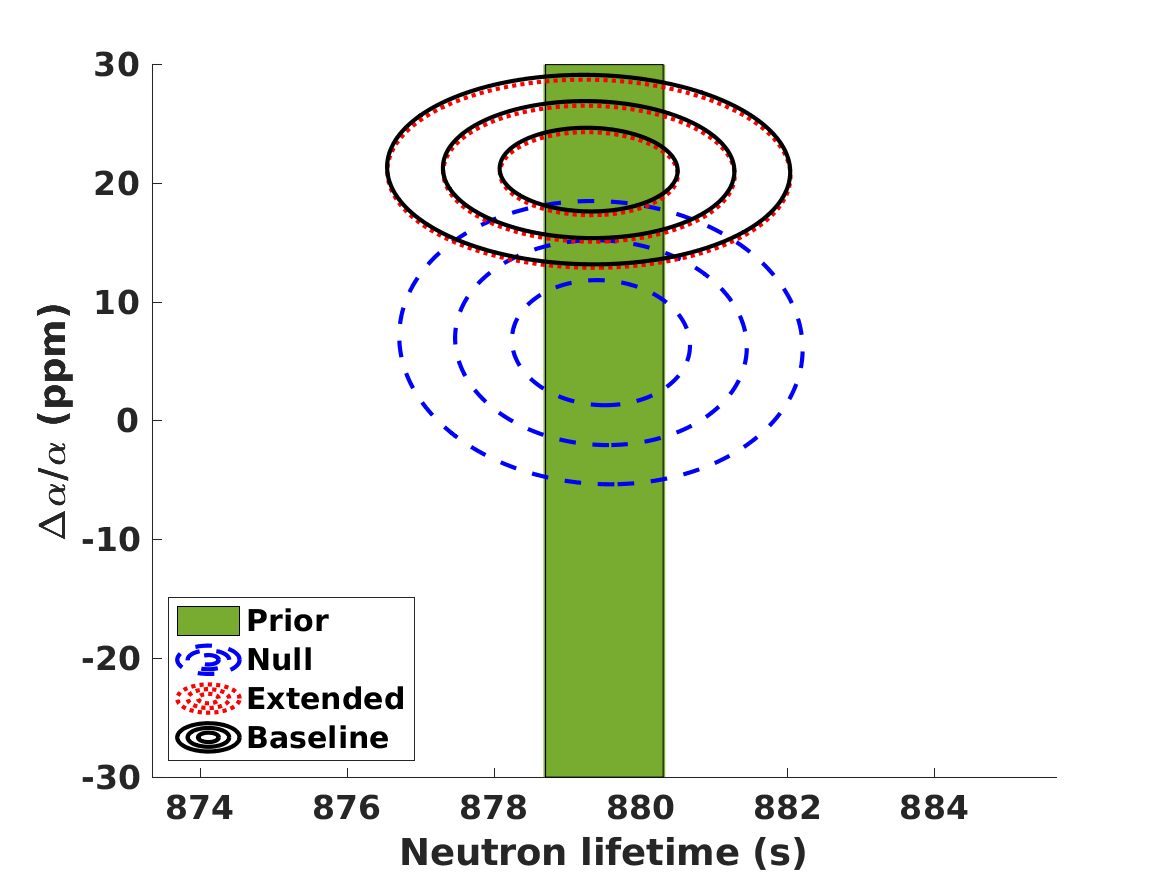}
\includegraphics[width=8cm]{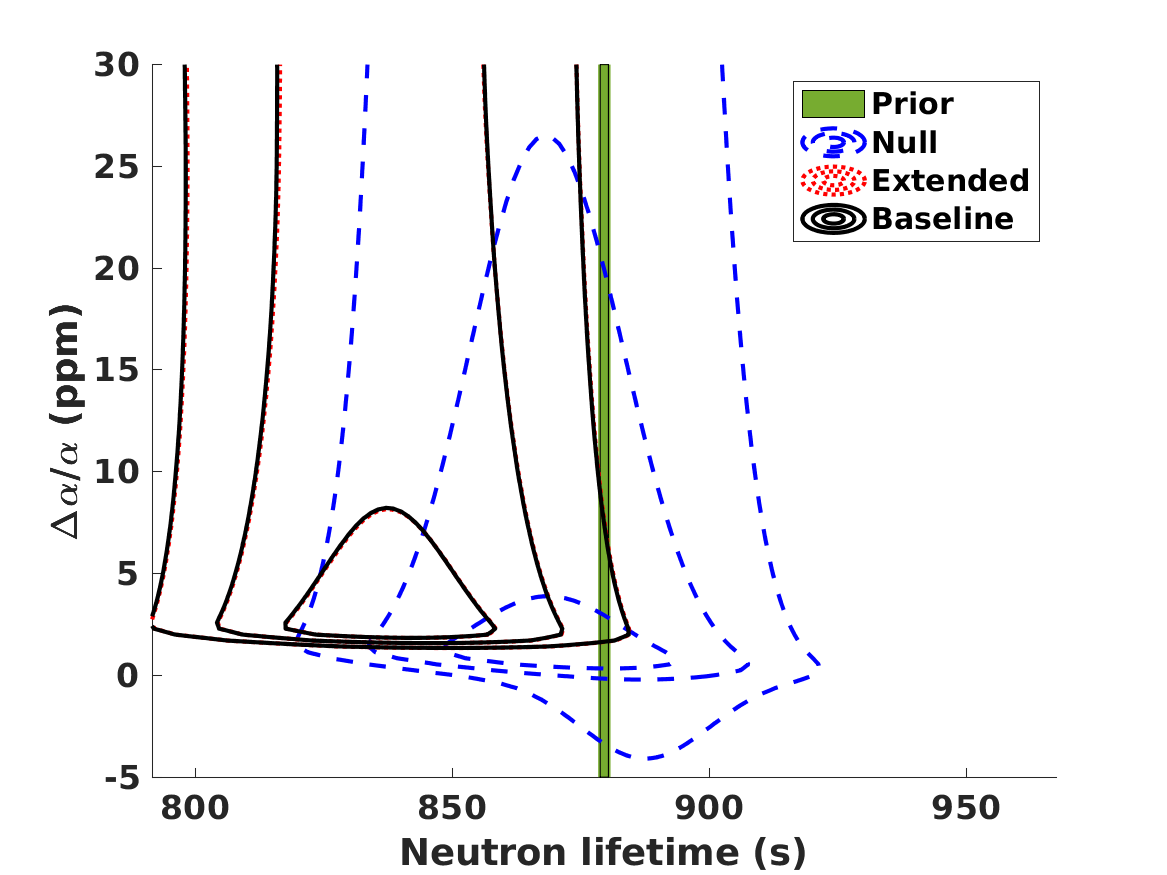}
\includegraphics[width=8cm]{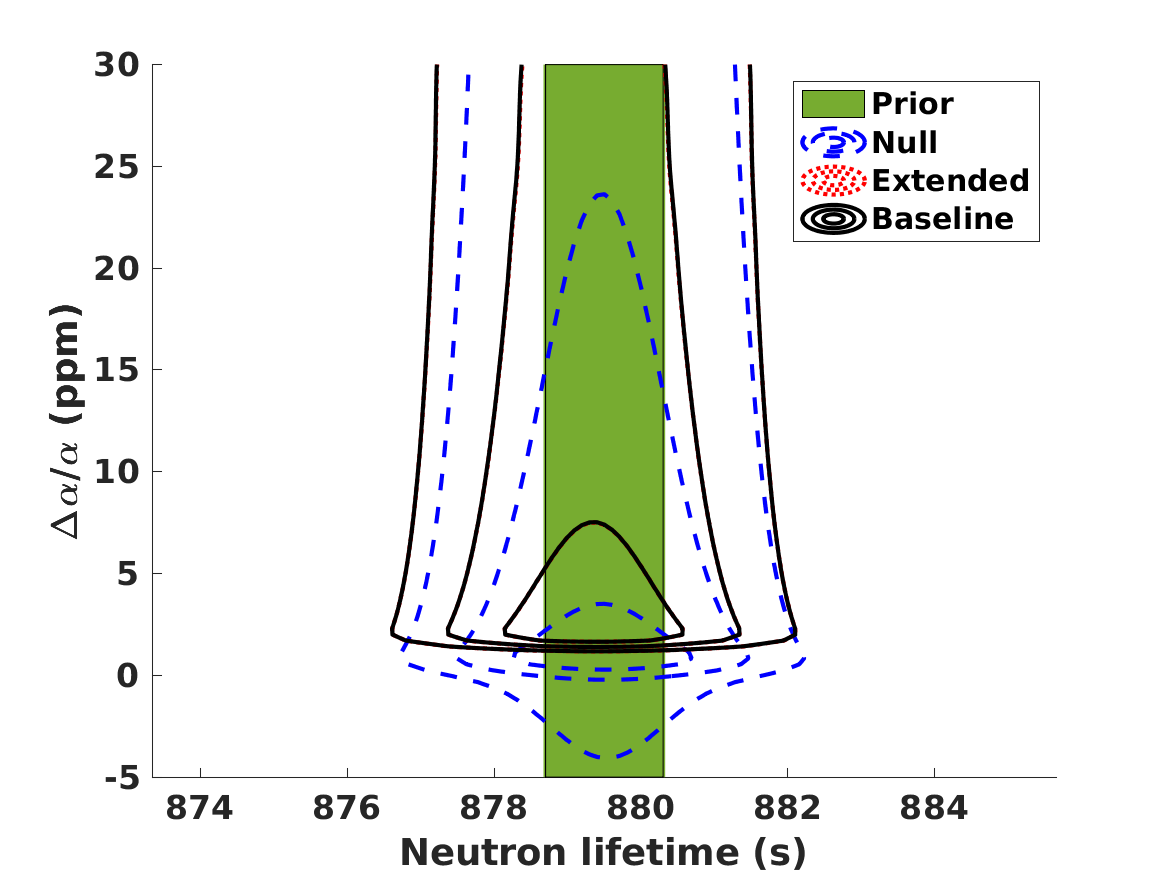}
\caption{Effect of the present-day neutron lifetime on the preferred value of $\alpha$ for unification, dilaton, and clocks cases (top, middle, and bottom panels, respectively). In the clocks case, the parameters $R$ and $S$ were marginalised. Left and right panels correspond to the cases without and with the experimental prior on the neutron lifetime discussed in the text; the one-sigma range corresponding to this prior is also shown for illustration purposes. It is important to notice that for each model the vertical axis is the same in both cases, while the horizontal one is not. The 68.3, 95.4, and 99.7 percent confidence levels are plotted throughout.}
\label{figure1}
\end{figure*}

We now repeat the previously discussed methodology for this case. Naturally, since these models could solve the lithium problem in the narrower parameter space (as shown in CM20), they will still do so in this wider parameter space.\ However, our goal is to assess how the best-fit models change due to possible correlations between the model parameters--in this case $\alpha$ and $\tau_n$,

The results of this analysis are depicted in Figure \ref{figure1}, and the derived constraints on $\alpha$ and $\tau_n$ are summarised in Table \ref{table5}. We note that in the latter, we report the best fit and the range of values within $\Delta\chi^2=4$ of it for each case. This would correspond to the $95.4\%$ confidence level for the case of a Gaussian posterior likelihood, which is the case (to a very good approximation) for the unification and dilaton models, but not for the clocks model.

%%%%%%%%%%%%%%%%%%%%
%%%%%%%%%%%%%%%%%%%%

Interestingly, if the neutron lifetime is allowed to be entirely free, the degeneracy between it and $\alpha$ is such that the  observational abundances would prefer a value of $\alpha$ that is somewhat larger than the one for the fixed $\tau_n$ (cf. the values obtained in CM20 and listed in Table \ref{table3}), while the preferred value of the neutron lifetime would be lower--even as low as 812 seconds in the case of the dilaton model. This is the case depicted on the left-hand side of  Figure \ref{figure1}. Although the associated error bars are large (close to 30 seconds, cf. Table \ref{table5}), these are obviously excluded by the aforementioned experiments at well over two standard deviations. Nevertheless, this is a useful alternative way to quantitatively express the lithium problem.

%%%%%%%%%%%%%%%%%%%%%%%%%%%%%%%%%%%%%%%%%%%%%%%%%%%%%%%%%%%%%%%%%%%%%%%%%%%%%%
\begin{table*}
\caption{Constraints on $\Delta\alpha/\alpha$ and $\tau_n$ for the unification, dilaton, and clocks models for the various choices of primordial abundances and priors used in the analysis. The listed values correspond to the best fit in each case and to the range of values within $\Delta\chi^2=4$ of it (corresponding to the $95.4\%$ confidence level for a Gaussian posterior likelihood).}
\label{table5}
\centering
\begin{tabular}{c | c c | c c | c c}
\hline
{}  & \multicolumn{2}{c}{Unification} & \multicolumn{2}{c}{Dilaton} & \multicolumn{2}{c}{Clocks} \\
Abundances  & $\Delta\alpha/\alpha$ (ppm) & $\tau_n$ (s) & $\Delta\alpha/\alpha$ (ppm) & $\tau_n$ (s) & $\Delta\alpha/\alpha$ (ppm) & $\tau_n$ (s)  \\
\hline
Baseline (No prior) & $13.5\pm3.0$ & $840.8\pm26.9$ & $26.2\pm5.0$ & $811.4\pm28.7$ & $2.3_{-1.6}^{+29.8}$ & $837.0\pm26.4$\\
Baseline (With prior) & $12.4\pm2.9$ & $879.3\pm1.6$ & $21.2\pm4.6$ & $879.3\pm1.6$ & $2.3_{-0.8}^{+17.2}$ & $879.4\pm1.6$\\
\hline
Null (No prior) & $5.1\pm4.2$ & $870.4\pm29.0$ & $11.8\pm10.2$ & $852.8\pm38.4$ & $1.1_{-1.0}^{+8.2}$ & $869.2\pm29.3$ \\
Null (With prior) & $4.6\pm3.8$ & $879.5\pm1.6$ & $6,6\pm6.9$ & $879.4\pm1.6$ & $0.8_{-0.9}^{+7.3}$ & $879.5\pm1.6$ \\
\hline
Extended (No prior) & $13.5\pm3.0$ & $841.0\pm27.0$ & $25.6\pm5.0$ & $812.7\pm28.7$ & $2.3_{-1.6}^{+25.7}$ & $837.0\pm26.4$ \\
Extended (With prior) & $12.4\pm4.9$ & $879.3\pm1.6$ & $20.8\pm4.6$ & $879.3\pm1.6$ & $2.3_{-0.8}^{+16.9}$ & $879.4\pm1.6$ \\
\hline
\end{tabular}
\end{table*}
%%%%%%%%%%%%%%%%%%%%%%%%%%%%%%%%%%%%%%%%%%%%%%%%%%%%%%%%%%%%%%%%%%%%%%%%%%%%%%

On the other hand, if we repeat the analysis using Eq. \ref{neutronpri} as a prior, then, as depicted on the right-hand side of  Figure \ref{figure1}, we approximately recover the best-fit values of $\alpha$ listed in Table \ref{table3}--although not exactly since the degeneracy between $\alpha$ and $\tau_n$ is not completely broken by the prior. Clearly, this degeneracy is also model-dependent--since it depends on the specific values of the parameters $R$ and $S$--and it is stronger for the dilaton case than for the unification one. This is the reason why in the latter case the best-fit values are almost unaffected.

Naturally, all the above considerations only apply if the observed lithium abundance is included in the analysis. If it is left out, then within the statistical uncertainties, the results are consistent with standard physics, as expected. We also confirm that the helium-3 abundance has no significant impact on the results. In any case, our analysis confirms that due to the experimental constraints, the inclusion of the neutron lifetime together with the other parameters in the perturbative analysis, as was done in CM20, is fully justified.

%%%%%%%%%%%%%%%%%%%%%%%%%%%%%%%%%%%%%%%%%%%%%%%%%%%%%%%%%%%%%%%
\section{Baryon-to-photon ratio and neutrino families}
\label{cosmo}

We now focus on degeneracies between the fine-structure constant and two cosmological parameters that are crucial for the BBN outcomes: the number of neutrino species and the baryon-to-photon ratio. On the other hand, the neutron lifetime will be included back in the main sensitivity coefficients for $\alpha$, as was done in CM20.

We note that in what follows, we express our constraints in terms of the number of neutrino species, $N_\nu$, rather than the effective number, which in the standard model has the value $N_{eff}=3.046$. The latter is conveniently constrained by several cosmological analyses. For example, a recent analysis combining cosmic microwave background and baryonic acoustic oscillations data \citep{Planck} leads to $N_{eff}=2.99\pm0.17$ or $N_{eff}=3.27\pm0.15$, depending on whether or non-local Hubble parameter data are also used. On the other hand, combining cosmic microwave background and BBN data \citep{Fields2} gives $N_\nu=2.86\pm0.15$. Moreover, the former is very tightly constrained by accelerator experiments. Specifically, the LEP measurement \citep{PDG}
\be
N_\nu=2.984\pm0.008\,
\ee
is used as a prior in our analysis.

For the baryon-to-photon ratio, the Planck satellite analysis finds \citep{Planck}
\be
\eta_{10}=6.104\pm0.058\,,
\ee
which we also use as a prior in our analysis. We note that for convenience, we have defined $\eta_{10}=\eta\times 10^{10}$. This could equivalently be expressed as a physical baryon density
\be
\omega_b=\Omega_bh^2=\frac{\eta_{10}}{274}=0.0223\pm0.0002\,.
\ee
Strictly speaking, we should note that this constraint on $\eta_{10}$ was obtained assuming the standard value of $\alpha$ during recombination. However, this is a safe assumption since there is no significant correlation between $\eta_{10}$ and $\alpha$ provided the cosmic microwave background temperature and polarisation are both used \citep{Planckalpha,Hart}.

In this case, our perturbative analysis has the following form
\begin{equation}
\frac{\Delta Y_i}{Y_i}=(x_i+y_iS+z_iR)\frac{\Delta\alpha}{\alpha}+w_i\frac{\Delta\eta}{\eta}+v_i\frac{\Delta N_\nu}{N_\nu}\,.
\end{equation}
Here the $\alpha$ sensitivity coefficients listed in Table \ref{table1} still apply, while those for $\eta$ (and equivalently $\eta_{10}$) and $N_\nu$ are listed in Table \ref{table6}. The latter have also been previously discussed in \citet{Pitrou} and also in \citet{Fields2}.

%%%%%%%%%%%%%%%%%%%%%%%%%%%%%%%%%%%%%%%%%%%%%%%%%%%%%%%%%%%%%%%%%%%%%%%%%%%%%%
\begin{table}
\caption{Sensitivity coefficients of BBN nuclide abundances on the free cosmological parameters of our phenomenological parametrisation, defined in the main text, and complementing those listed in Table \ref{table1}.}
\label{table6}
\centering
\begin{tabular}{| c | c c c c |}
\hline
$C_{ij}$ & D & ${}^3$He & ${}^4$He & ${}^7$Li \\
\hline
$w_i$ & -1.65 & -0.567 & +0.039 & +2.08 \\
$v_i$ & +0.409 & +0.136 & +0.164 & -0.277 \\
\hline
\end{tabular}
\end{table}
%%%%%%%%%%%%%%%%%%%%%%%%%%%%%%%%%%%%%%%%%%%%%%%%%%%%%%%%%%%%%%%%%%%%%%%%%%%%%%

We now proceed with the analysis along the same lines as in the previous section. We now have a parameter space comprising $\alpha$, $N_\nu$, and $\eta_{10}$, together with the phenomenological parameters $R$ and $S$ in the case of the clocks model. As before, we consider the cases without and with priors, and in the latter case we simultaneously use the LEP prior for $N_\nu$ and the Planck prior for $\eta_{10}$. For the unification and dilaton models, we report, as before, the best fit and the range of values within $\Delta\chi^2=4$ of it. On the other hand, for the clocks model, given the wider (five-dimensional) parameter space and the numerical difficulties in accurately sampling it, we instead report the best fit and the range of values within $\Delta\chi^2=1$ of it.

%%%%%%%%%%%%%%%%%%%%%%%%%%%%%%%%%%%%%%%%%%%%%%%%%%%%%%%%%%%%%%%%%%%%%%%%%%%%%%
\begin{table*}
\caption{Constraints on $\Delta\alpha/\alpha$, $N_\nu$, and $\eta_{10}$ for the unification and dilaton models, for the various choices of primordial abundances used in the analysis. The listed values correspond to the best fit in each case and to the range of values within $\Delta\chi^2=4$ of it (corresponding to the $95.4\%$ confidence level for a Gaussian posterior likelihood).}
\label{table7}
\centering
\begin{tabular}{c | c c c | c c c}
\hline
{}  & \multicolumn{3}{c}{Unification} & \multicolumn{3}{c}{Dilaton} \\
Abundances  & $\Delta\alpha/\alpha$ (ppm) & $N_\nu$ & $\eta_{10}$ & $\Delta\alpha/\alpha$ (ppm) & $N_\nu$ & $\eta_{10}$ \\
\hline
Baseline (No priors) & $30.2\pm6.6$ & $2.90\pm0.42$ & $6.79\pm0.35$ & $34.0\pm7.5$ & $1.86\pm0.43$ & $5.92\pm0.23$ \\
Baseline (With priors) & $16.1\pm3.9$ & $2.98\pm0.02$ & $6.25\pm0.10$ & $22.6\pm5.5$ & $2.98\pm0.02$ & $6.18\pm0.10$\\
\hline
Null (No priors) & $-2.0\pm42.2$ & $2.86\pm0.42$ & $5.85\pm1.54$ & $-1.0\pm33.6$ & $2.88\pm1.06$ & $5.91\pm0.23$ \\
Null (With priors) & $4.6\pm5.5$ & $2.98\pm0.02$ & $6.10\pm0.10$ & $3.7\pm8.4$ & $2.98\pm0.02$ & $6.06\pm0.10$ \\
\hline
Extended (No priors) & $29.8\pm6.6$ & $2.90\pm0.42$ & $6.77\pm0.35$ & $34.2\pm7.3$ & $1.89\pm0.43$ & $5.90\pm0.23$  \\
Extended (With priors) & $16.0\pm3.9$ & $2.98\pm0.02$ & $6.25\pm0.10$ & $22.1\pm5.5$ & $2.98\pm0.02$ & $6.18\pm0.10$ \\
\hline
\end{tabular}
\end{table*}
%%%%%%%%%%%%%%%%%%%%%%%%%%%%%%%%%%%%%%%%%%%%%%%%%%%%%%%%%%%%%%%%%%%%%%%%%%%%%%

\begin{figure*}
\centering
\includegraphics[width=8cm]{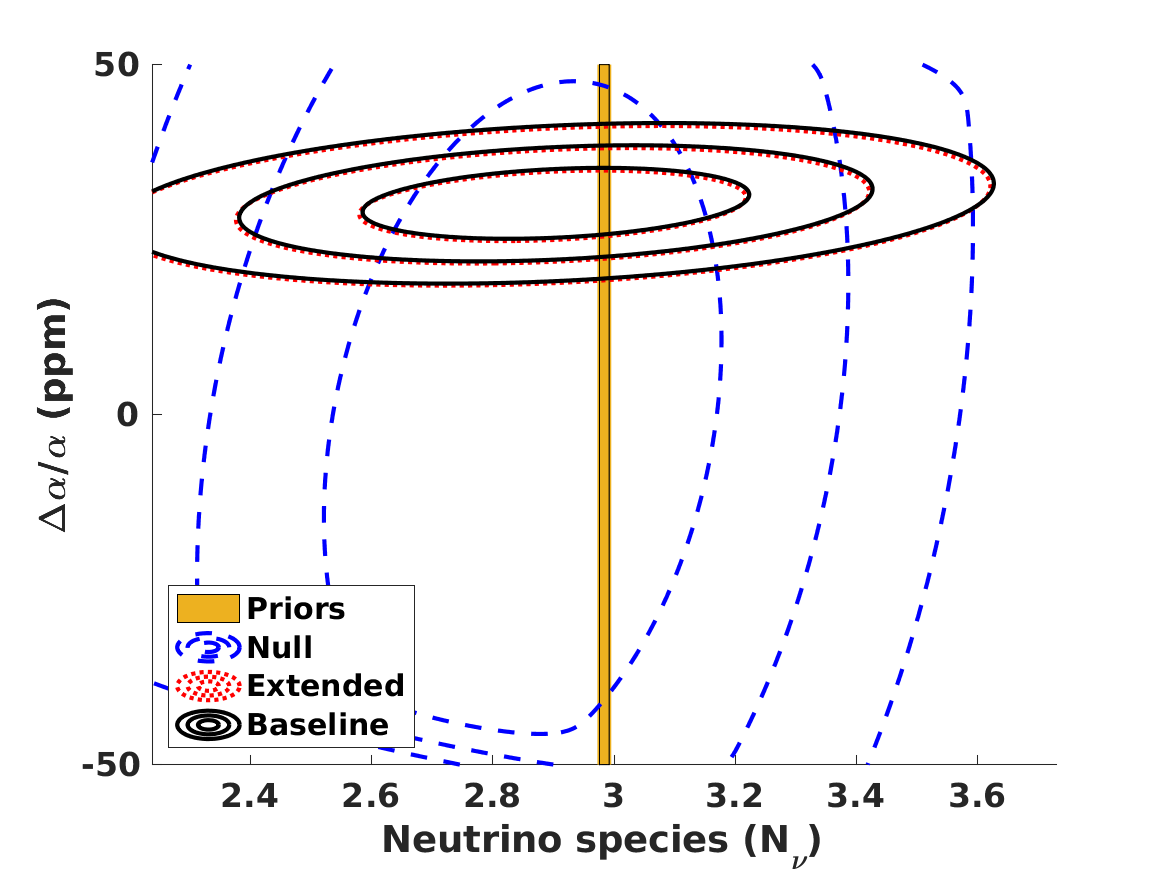}
\includegraphics[width=8cm]{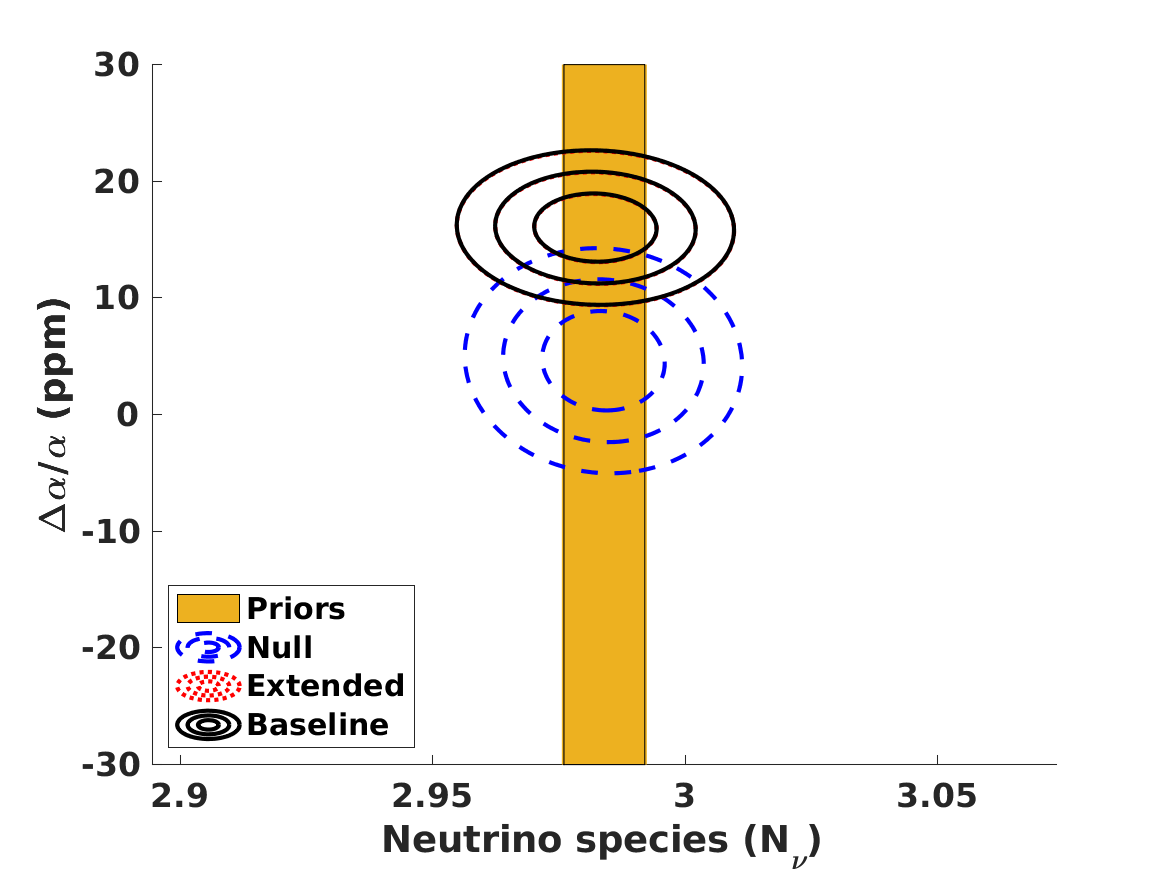}
\includegraphics[width=8cm]{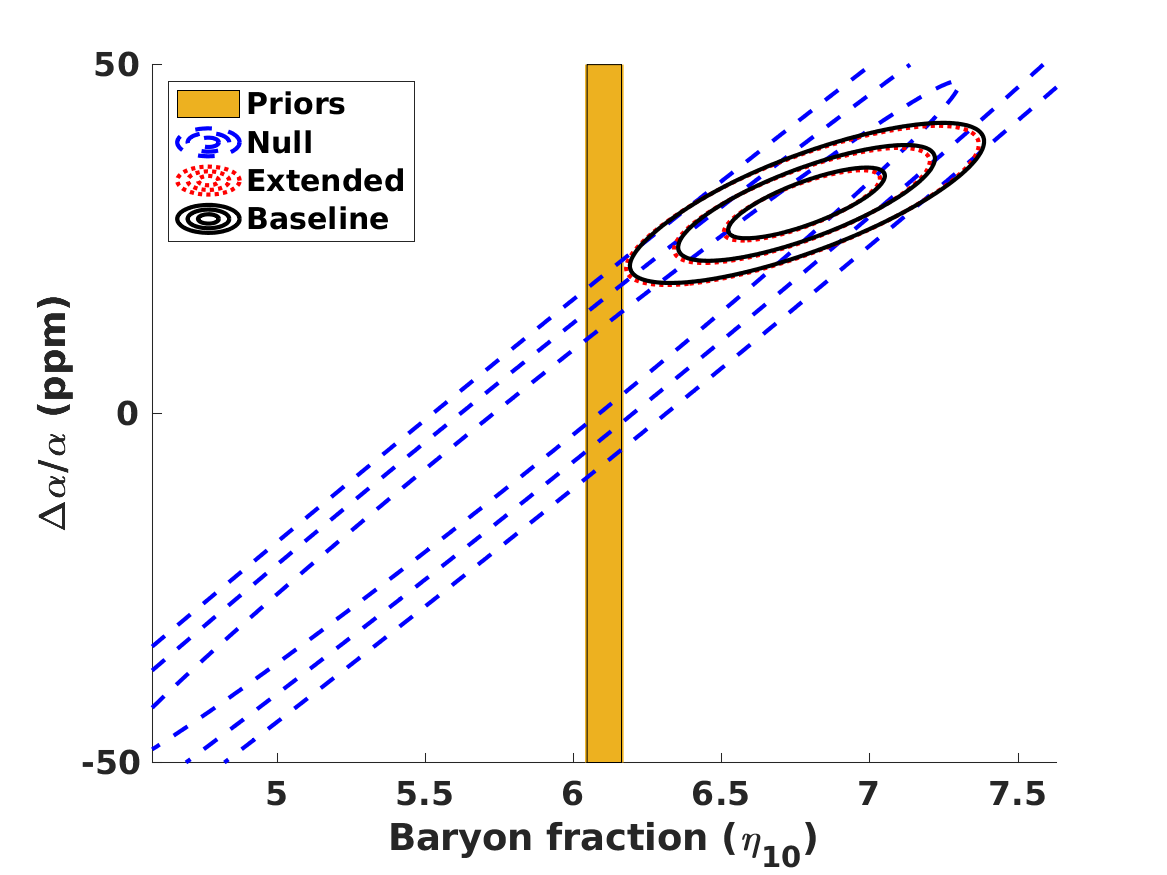}
\includegraphics[width=8cm]{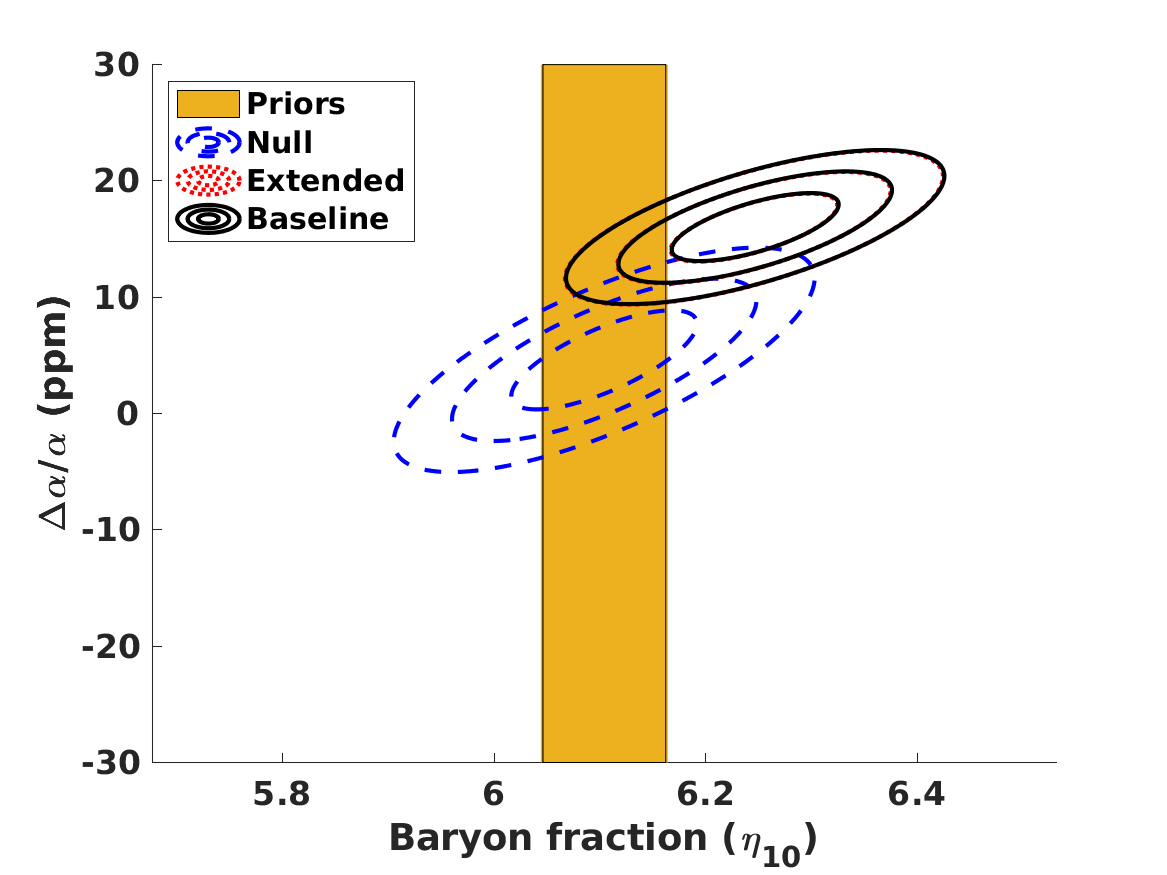}
\includegraphics[width=8cm]{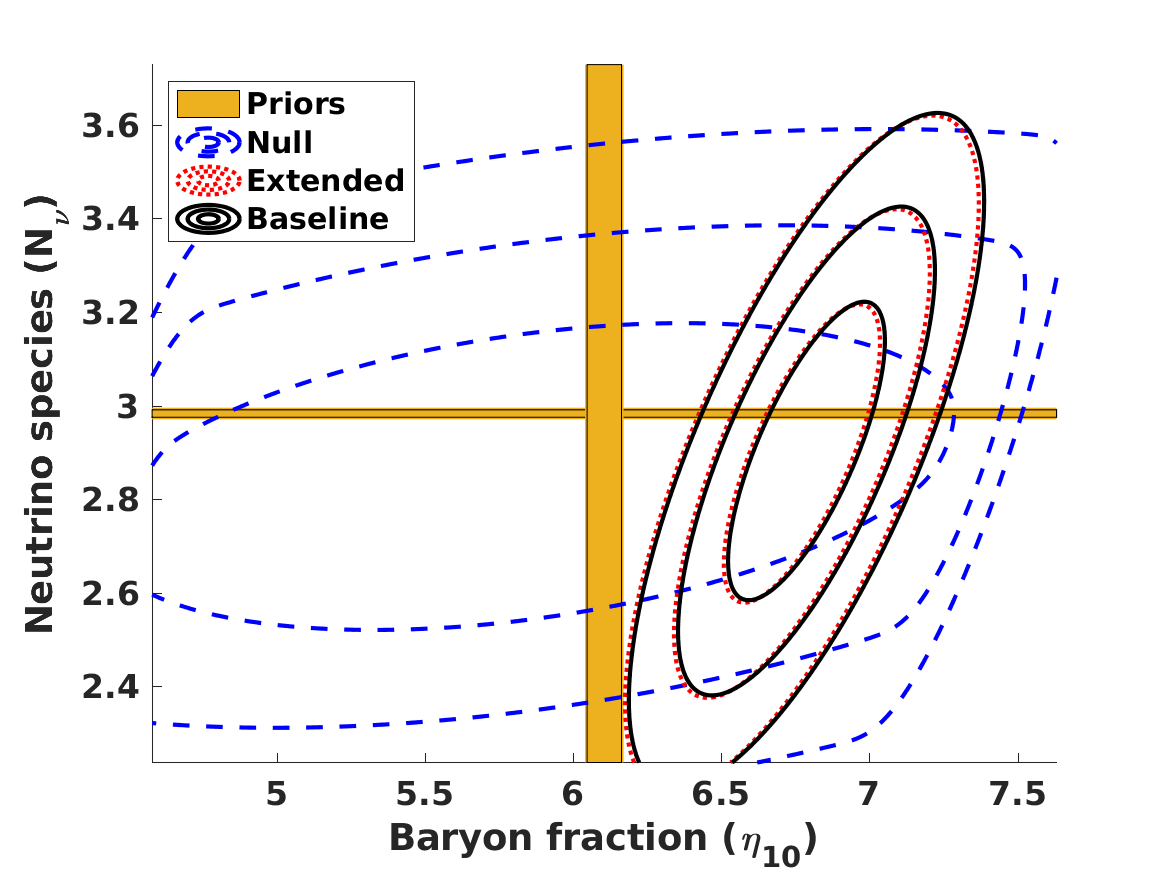}
\includegraphics[width=8cm]{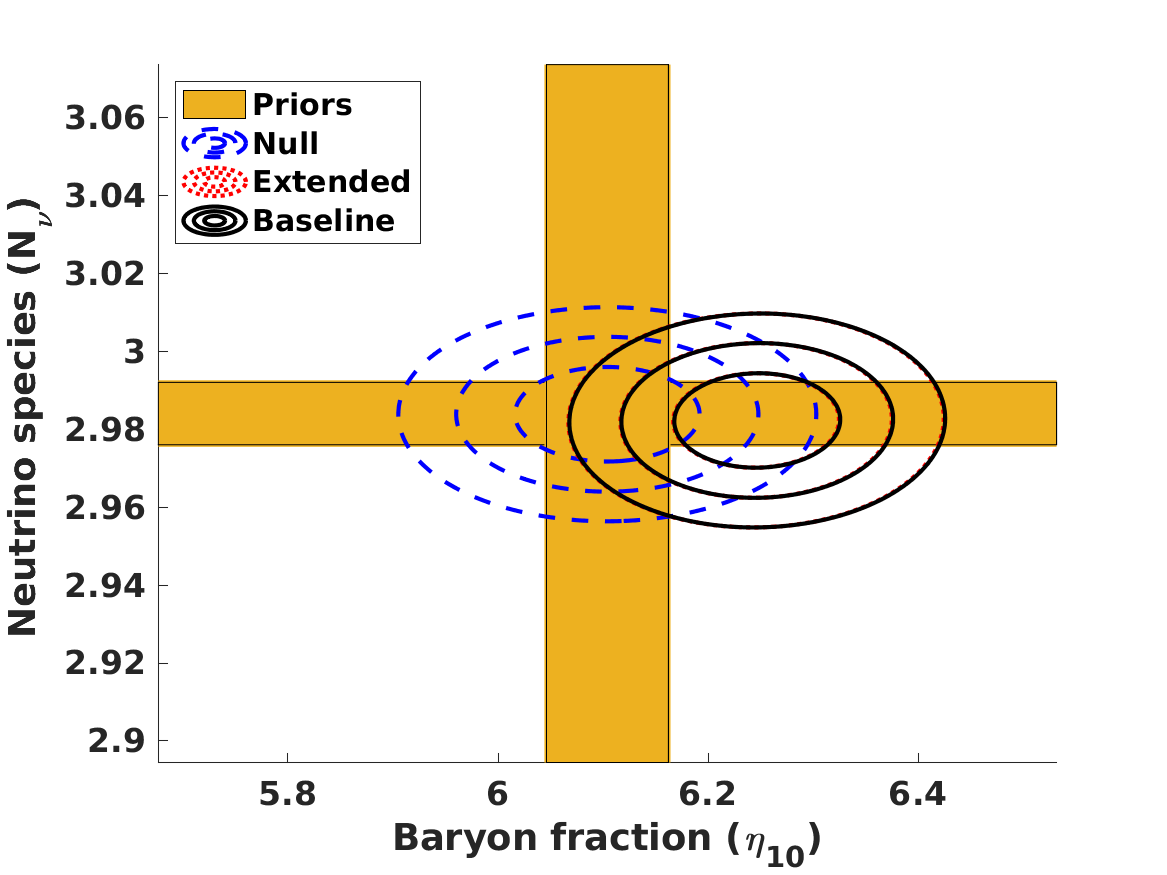}
\caption{Effect of the baryon-to-photon ratio and number of light neutrino species on the preferred value of $\alpha$ for the unification model. Left and right panels correspond to the cases without and with the two priors discussed in the text; the one-sigma range corresponding to these priors is also shown for illustration purposes. It is important to notice the different axis ranges in the left and right panels for each case. In each panel, the parameters that are not displayed were marginalised. The 68.3, 95.4, and 99.7 percent confidence levels are plotted throughout.}
\label{figure2}
\end{figure*}

 We start with the unification model, for which the results of our analysis can be found in Table \ref{table7} and in Figure \ref{figure2}. As in the case of the neutron lifetime, if no priors are used, we find that the degeneracies between the model parameters are such that when the lithium abundance is included, the best-fit values of $\alpha$ increase with respect to the ones in Table \ref{table3}. Interestingly though, this degeneracy does not impact the two other cosmological parameters in the same way: $N_\nu$ is unaffected (its best-fit value being fully compatible with the standard one), while $\eta_{10}$ is drastically affected with the preferred value being $\eta_{10}\sim6.8$; in other words, the BBN data alone would favour a much larger value of $\eta_{10}$ than is allowed by the Planck data.
 
 The above trend still persists when the LEP and Planck priors are added to the analysis: The values of $\alpha$ become comparable to (but still slightly larger than) those of Table \ref{table3}, $N_\nu$ exactly matches the prior value (with the error bar becoming slightly larger), while the preferred baryon-to-photon ratio increases to $\eta_{10}=6,25\pm0.10$ at the $95.4\%$ confidence level. Thus in the context of the unification model, the lithium problem could be recast as a baryon-to-photon ratio problem.
 
 We note that, even in the case of the null analysis, opening up the parameter space by allowing $\alpha$ to vary, and then marginalising it, significantly weakens the constraints on the $\eta_{10}-N_\nu$ plane even in the null case, as can be seen in the bottom left panel of Figure \ref{figure2}. This is to be expected, given the correlations between the three parameters, and can be mitigated by including priors on $\eta_{10}$ and/or $N_\nu$ in the analysis, which is the case depicted in the bottom right panel of the same figure.
 
%%%%%%%%%%%%%%%%%%%%
%%%%%%%%%%%%%%%%%%%%

\begin{figure*}
\centering
\includegraphics[width=8cm]{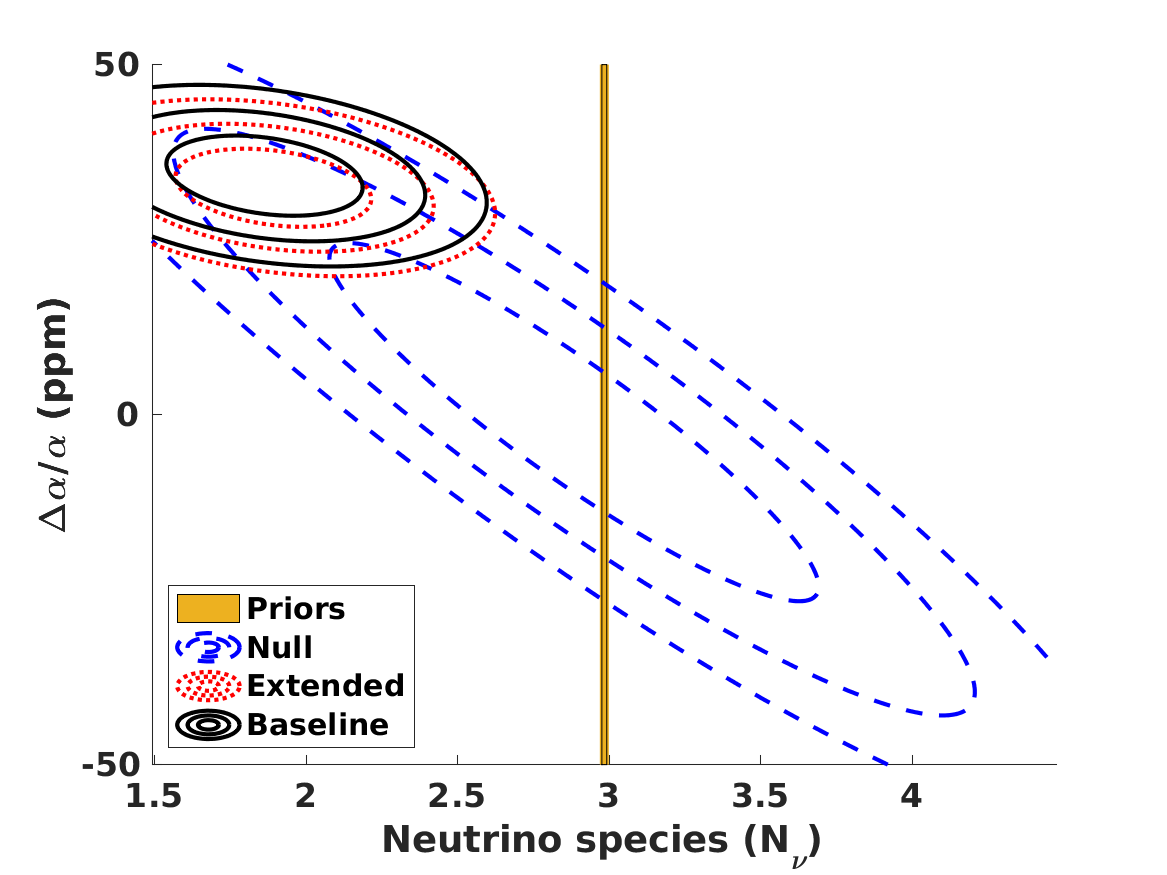}
\includegraphics[width=8cm]{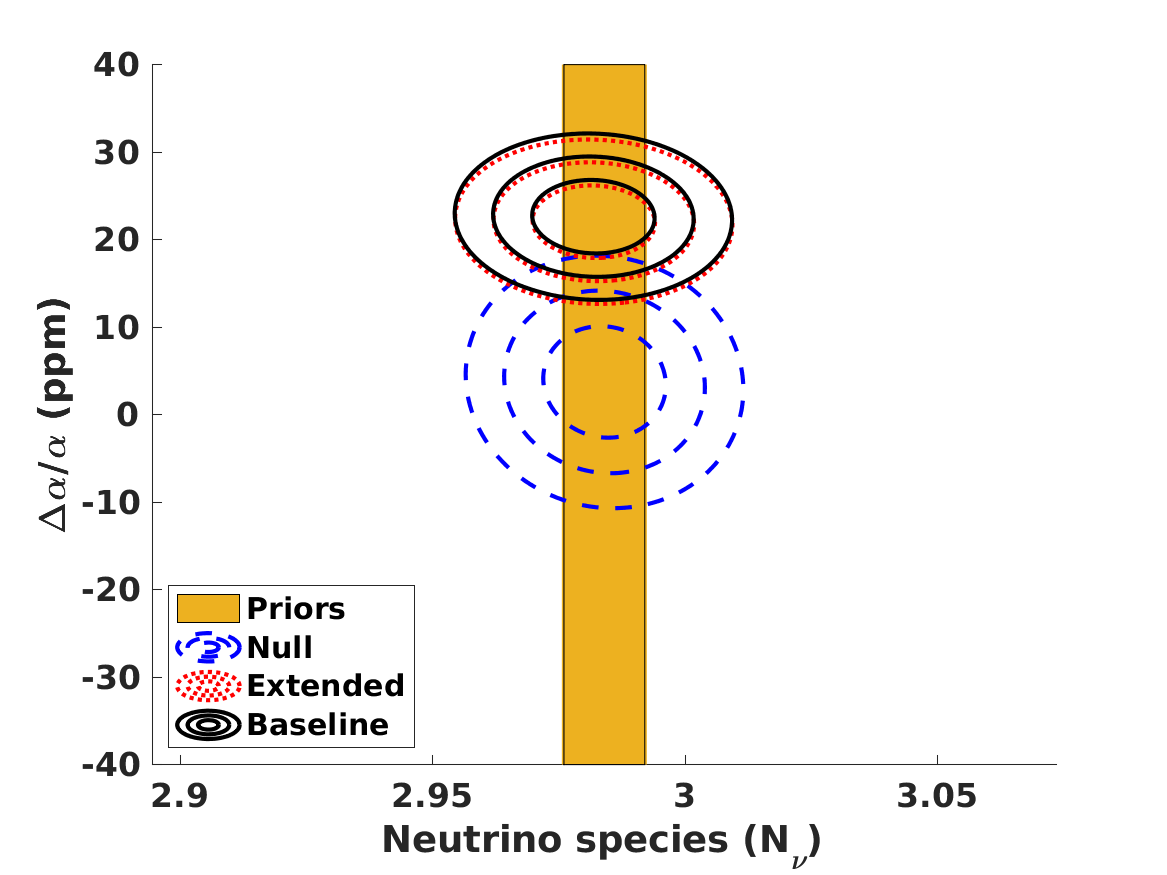}
\includegraphics[width=8cm]{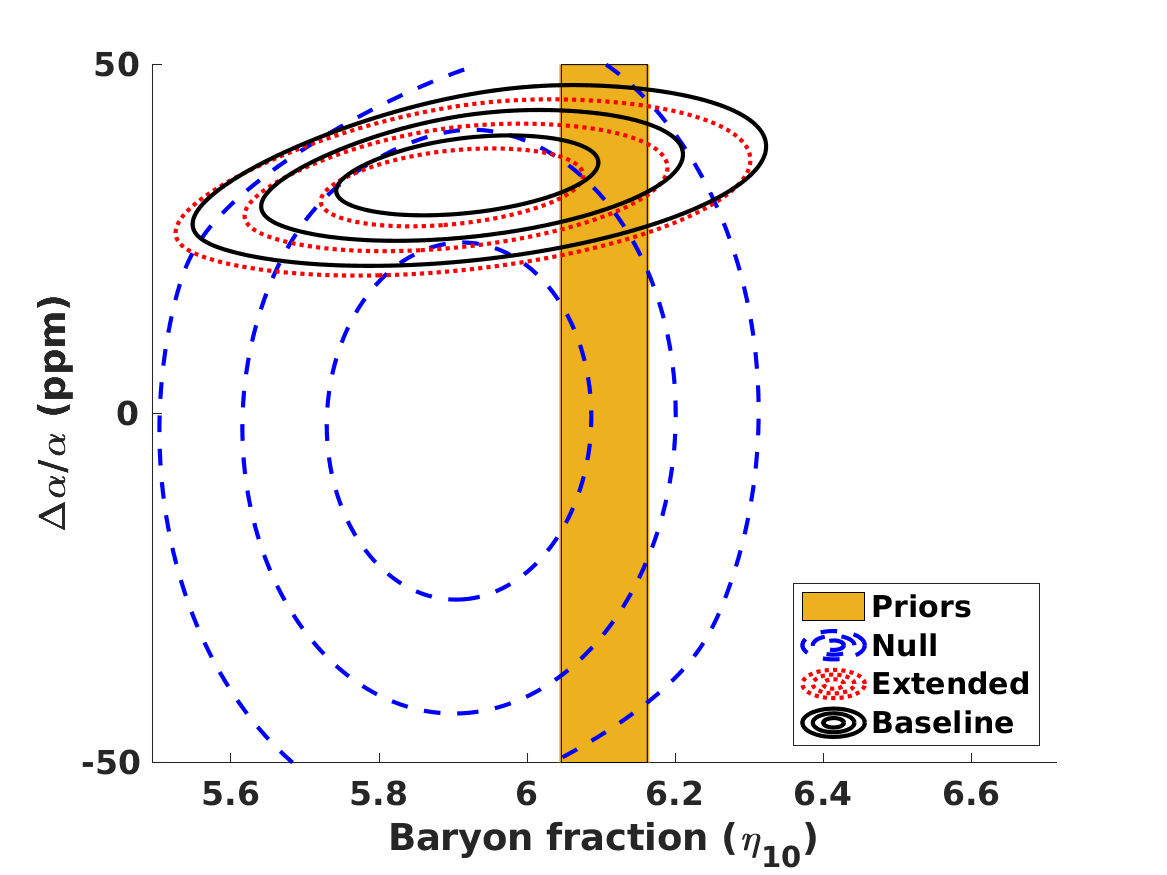}
\includegraphics[width=8cm]{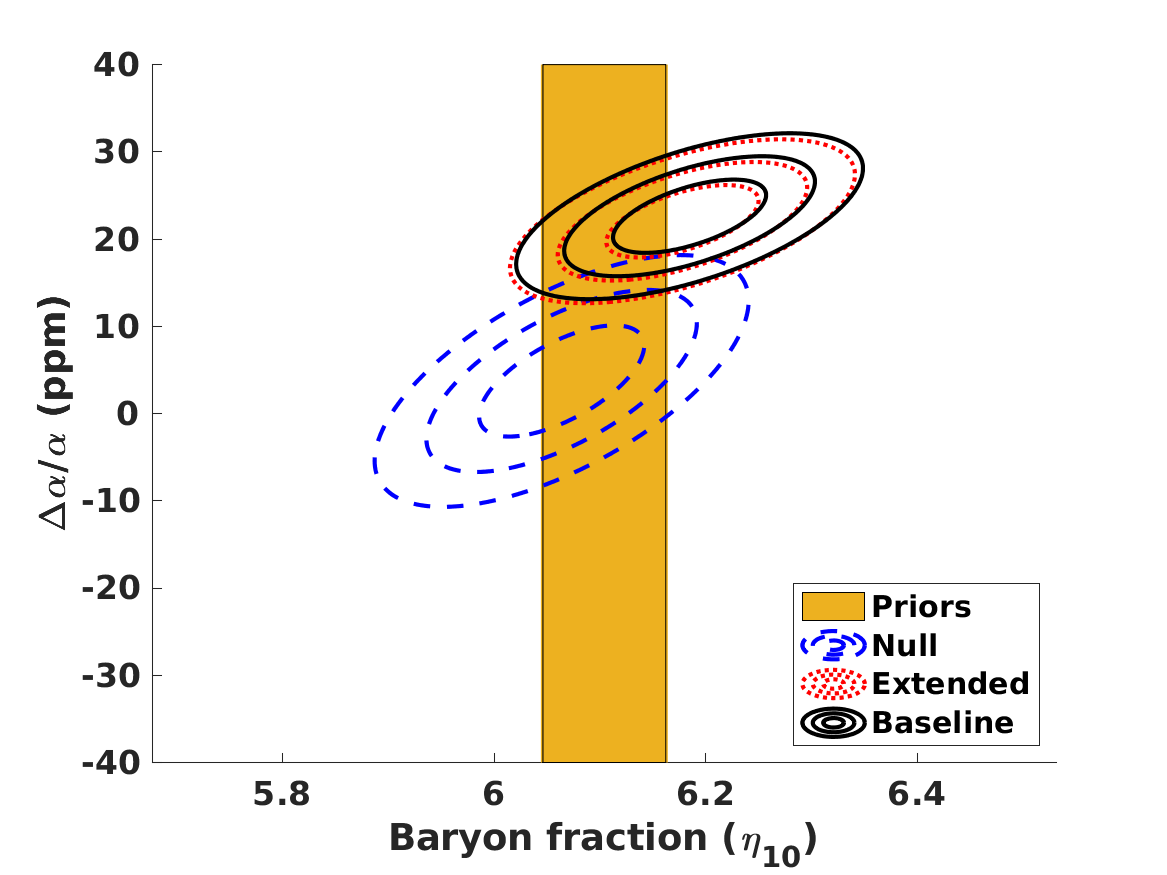}
\includegraphics[width=8cm]{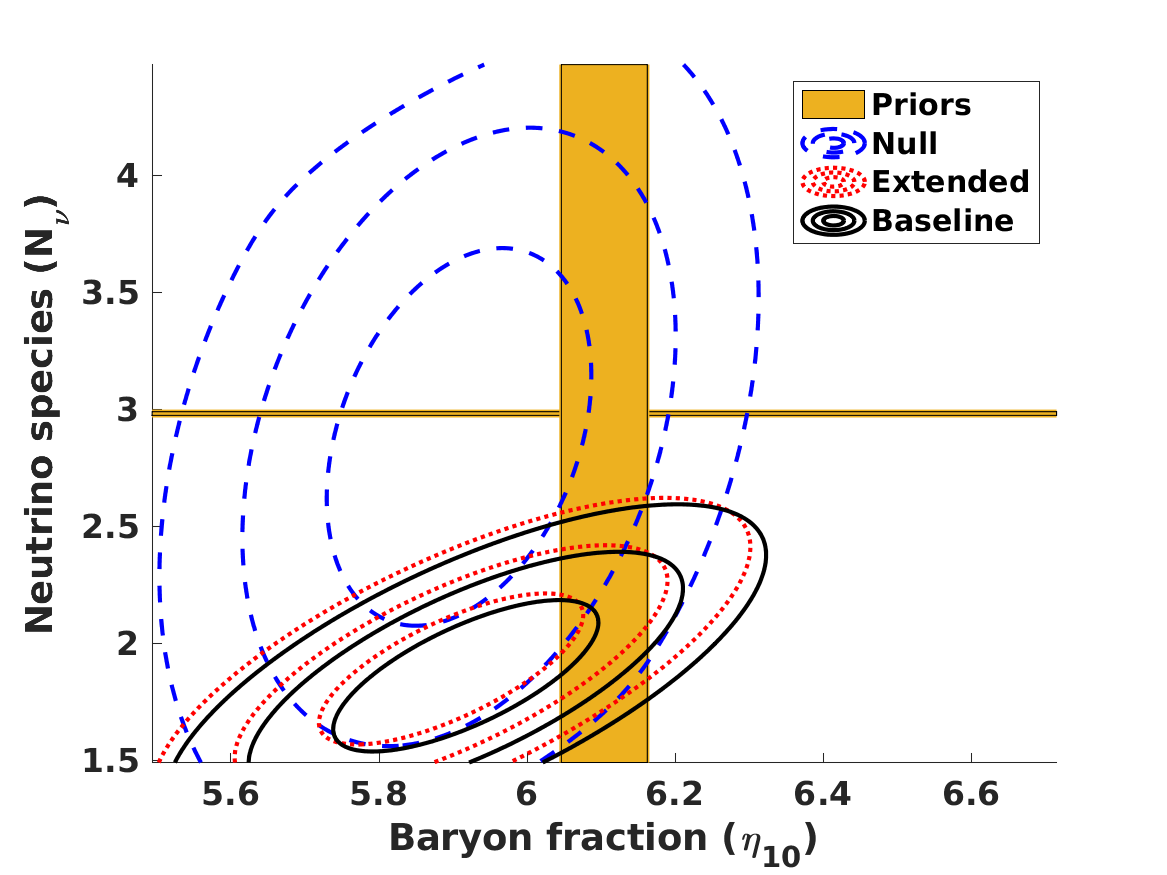}
\includegraphics[width=8cm]{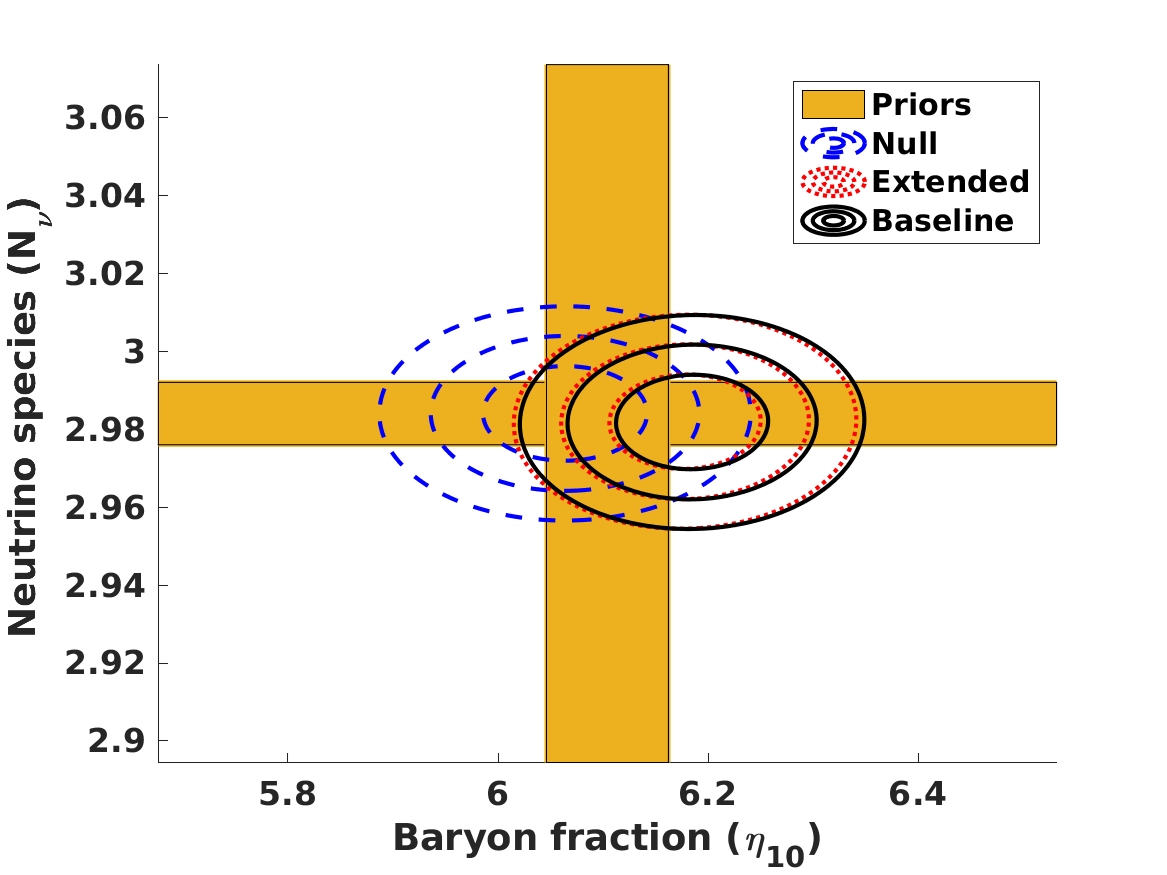}
\caption{Effect of the baryon-to-photon ratio and number of light neutrino species on the preferred value of $\alpha$ for the dilaton model. Left and right panels correspond to the cases without and with the two priors discussed in the text; the one-sigma range corresponding to these priors is also shown for illustration purposes. It is important to notice the different axis ranges in the left and right panels for each case. In each panel, the parameters that are not displayed were marginalised. The 68.3, 95.4, and 99.7 percent confidence levels are plotted throughout.}
\label{figure3}
\end{figure*}

The fact that the above analysis is model-dependent can be easily seen by contrasting the results for the unification model with the analogous results for the dilaton model, which can be found in Table \ref{table7} and in Figure \ref{figure3}. In this case, if no priors are used, the best-fit value of $\alpha$ also increases, but here the degeneracy is mainly with the number of neutrino species for which a value of $N_\nu\sim1.9$ would be preferred, which would be in obvious conflict with the LEP result and even other cosmological constraints on $N_{eff}$. On the other hand, the effect on $\eta_{10}$ is much smaller: While the best-fit value decreases slightly, it is still compatible with the standard values.

The inclusion of the LEP and Planck priors increases the preferred values of the number of neutrino species and the baryon-to-photon ratio, while simultaneously decreasing the preferred value of $\alpha$. Clearly the LEP prior dominates the analysis, and the best-fit value of $\eta_{10}$ is slightly larger than the standard one, though fully compatible with it given the increased statistical uncertainties in this larger parameter space. In any case, in the context of the dilaton model, the lithium problem could be recast as a neutrino species problem.

%%%%%%%%%%%%%%%%%%%%
%%%%%%%%%%%%%%%%%%%%

\begin{figure*}
\centering
\includegraphics[width=8cm]{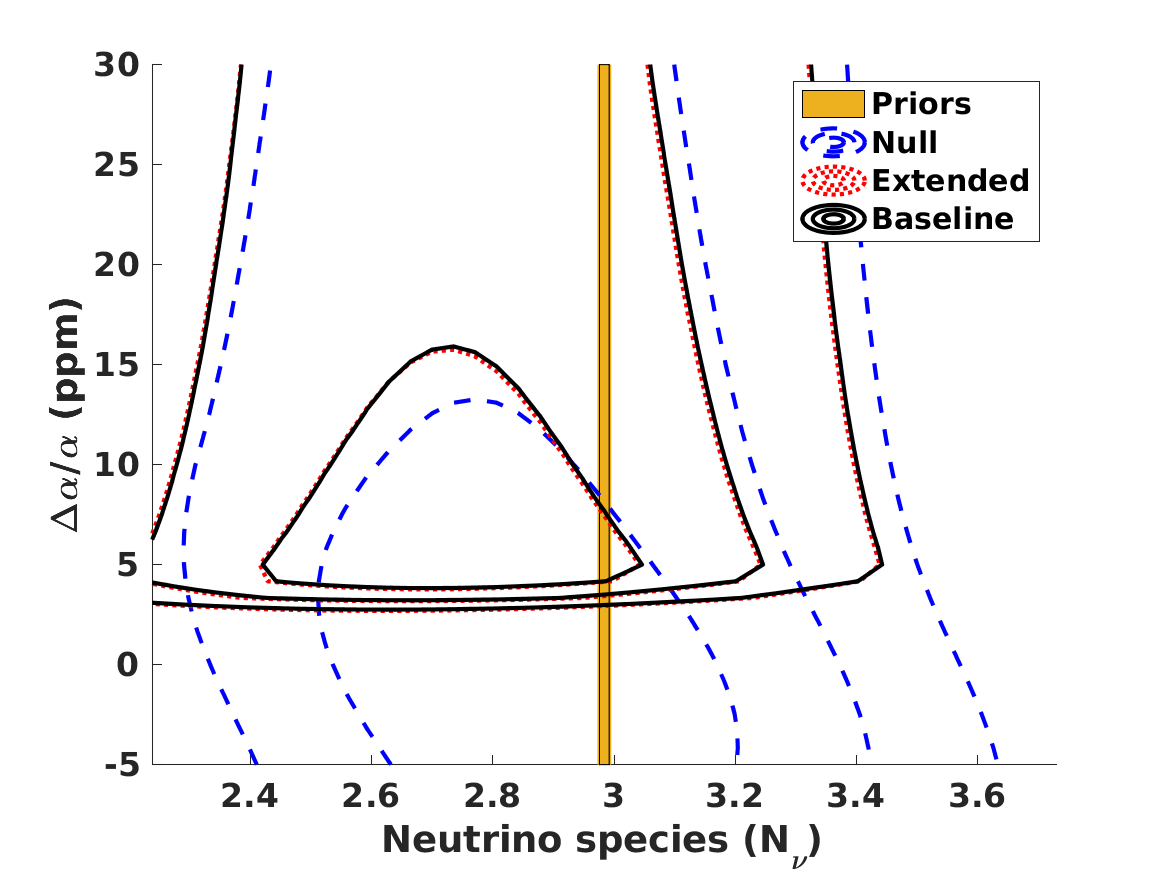}
\includegraphics[width=8cm]{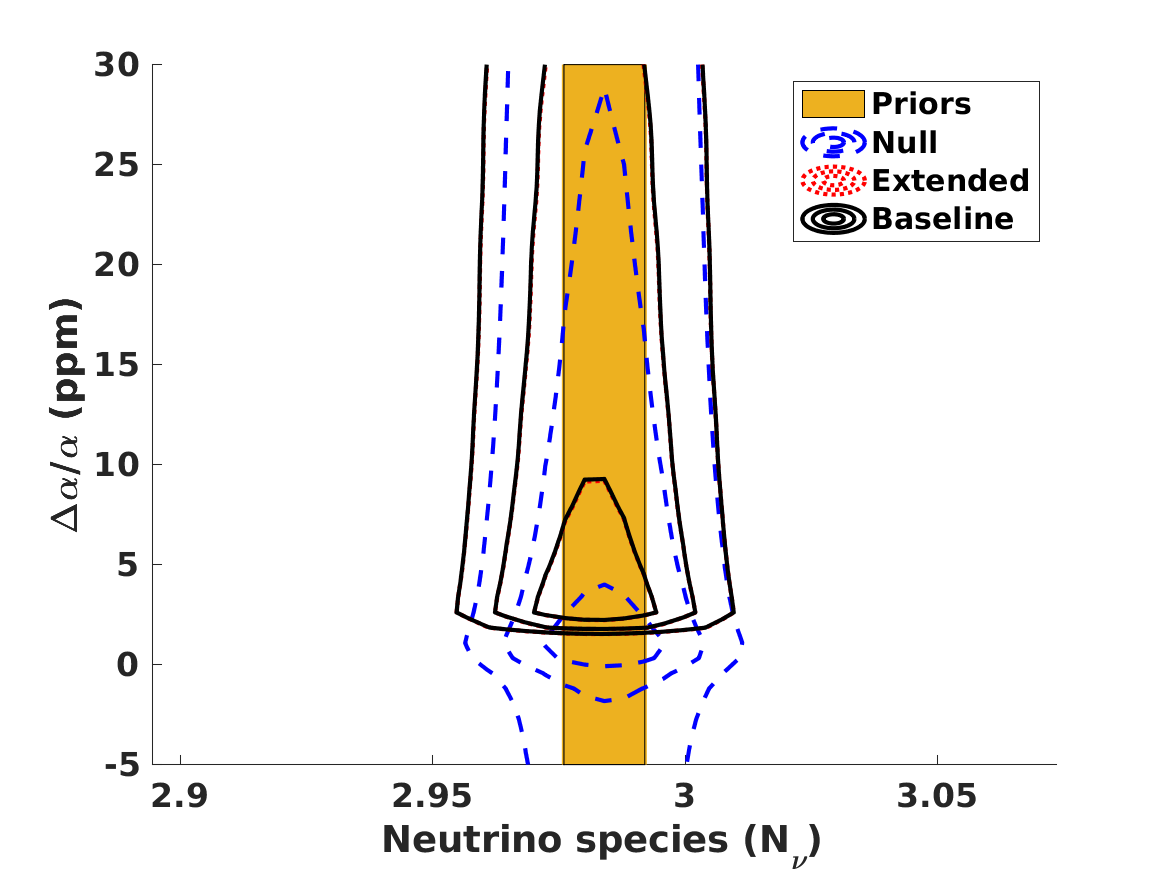}
\includegraphics[width=8cm]{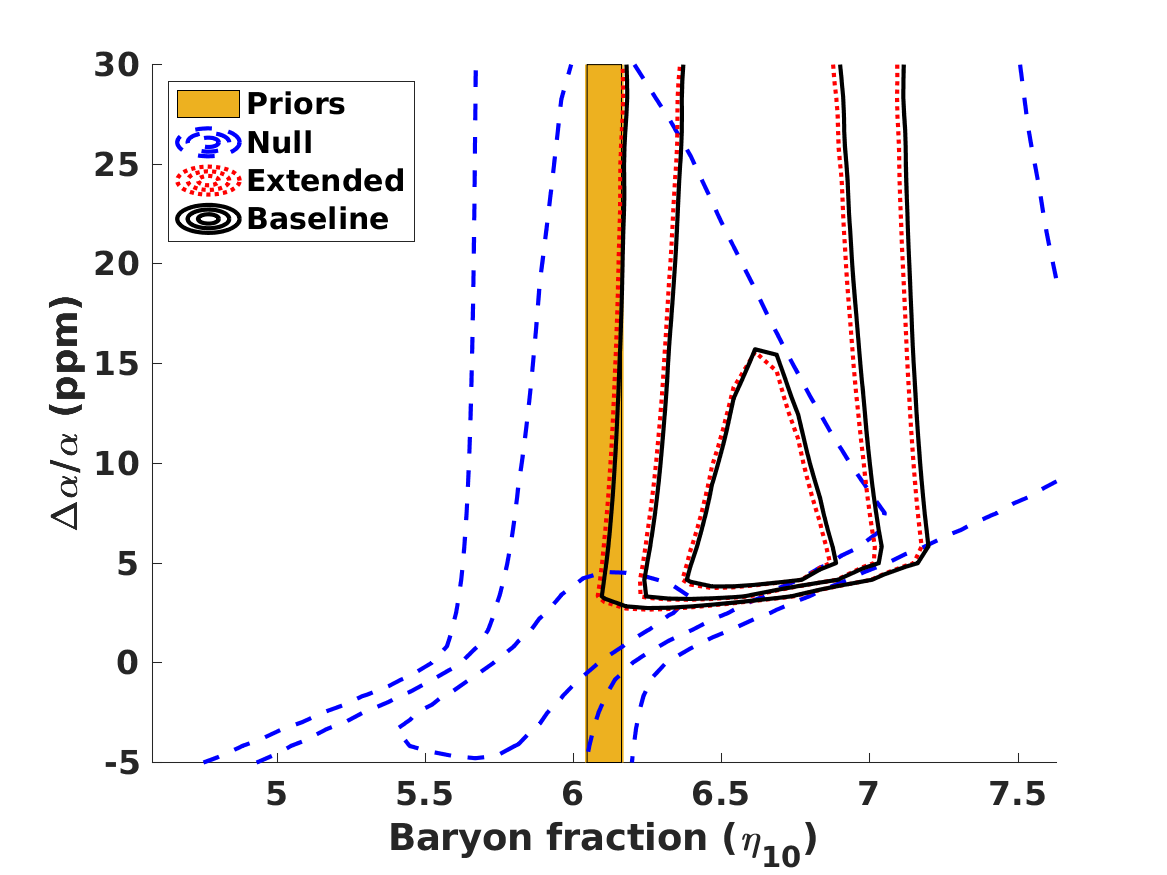}
\includegraphics[width=8cm]{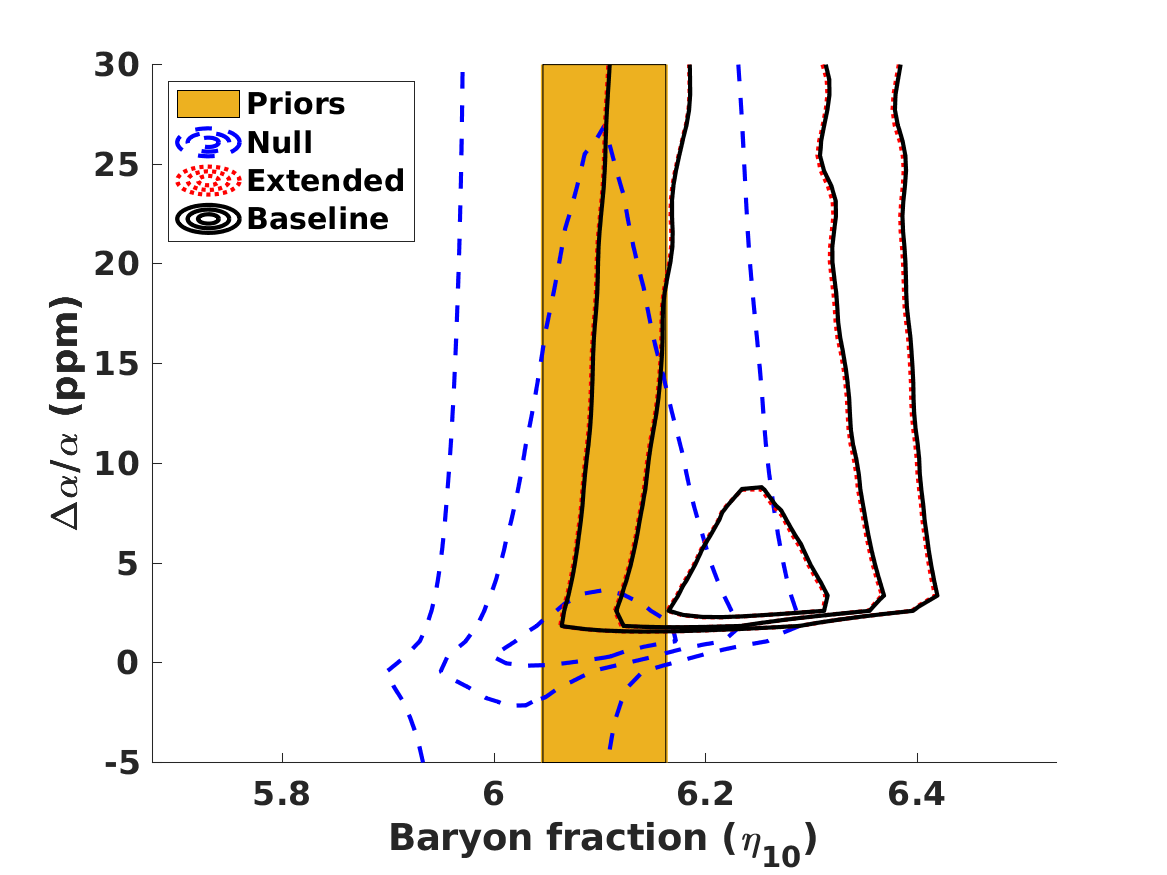}
\includegraphics[width=8cm]{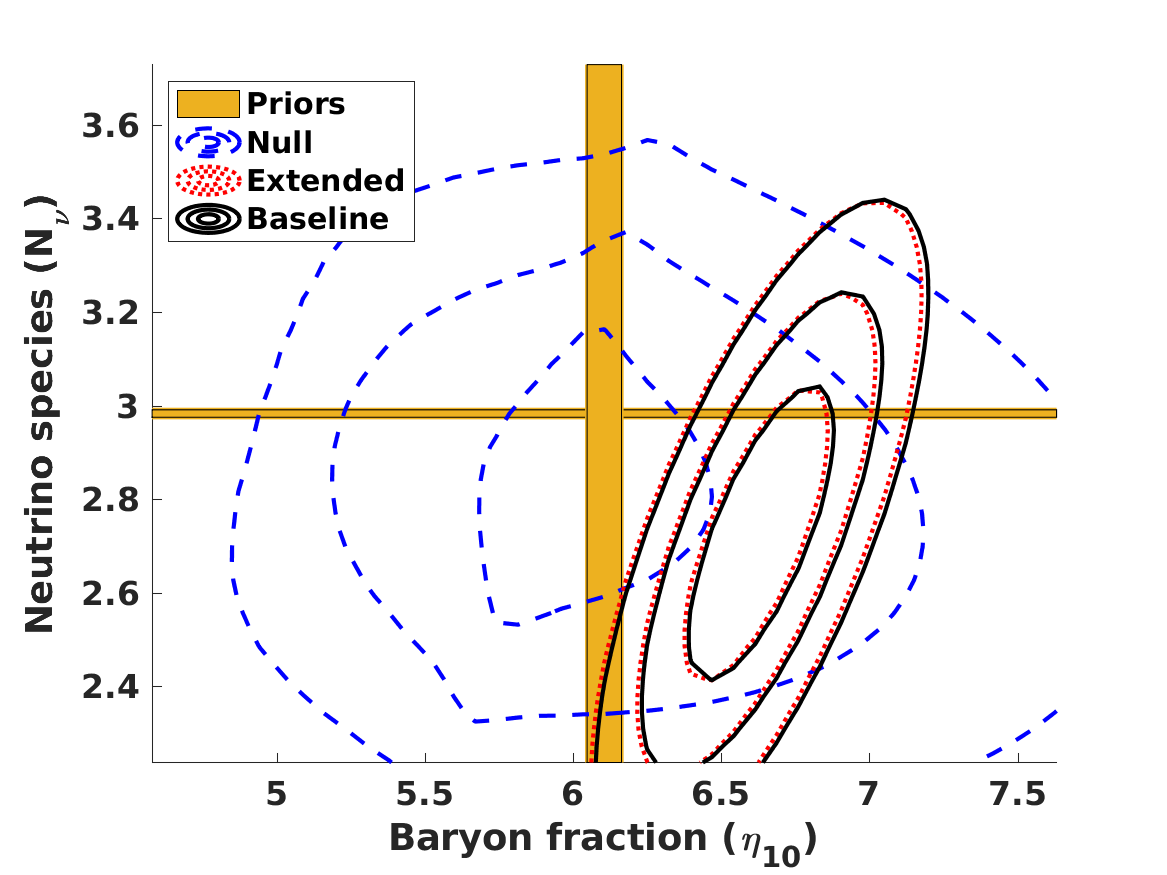}
\includegraphics[width=8cm]{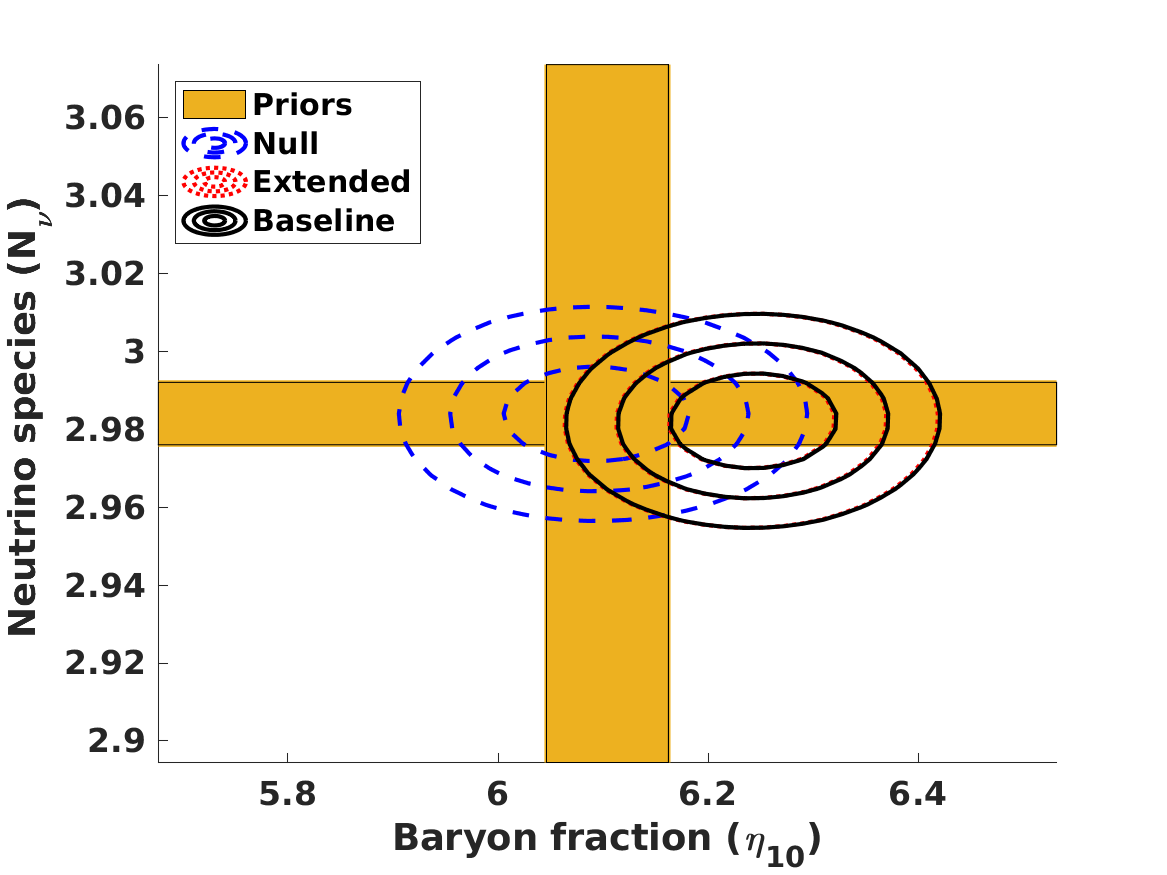}
\caption{Effect of the baryon-to-photon ratio and number of light neutrino species on the preferred value of $\alpha$ for the clocks model. Left and right panels correspond to the cases without and with the two priors discussed in the text; the one-sigma range corresponding to these priors is also shown for illustration purposes. It is important to notice the different axis ranges in the left and right panels for each case. In each panel, the parameters that are not displayed were marginalised. The 68.3, 95.4, and 99.7 percent confidence levels are plotted throughout.}
\label{figure4}
\end{figure*}

For the more generic clocks model, the analogous results can be found in Table \ref{table8} and in Figure \ref{figure4}. Since the parameters $R$ and $S$ are now allowed to vary and then be marginalised (as opposed to being kept fixed), the results are, in a loose sense, an average of those for the various specific models. In this case (and with our previously discussed choices of priors for $R$ and $S$), the result is closer to that of the unification model than to that of the dilaton model. Specifically, without the LEP and Planck priors, a significantly larger value of $\eta_{10}\sim6.6$ is preferred, while that of $N_\nu$ slightly decreases but is still compatible with the standard value, given the statistical uncertainties. In this case a relative variation of $\alpha$ at the 5 ppm level is preferred, with a very asymmetric (and obviously non-Gaussian) posterior likelihood.

Including the LEP and Planck priors, the former carries most of the statistical weight, it and decreases both the best-fit value of the relative variation of $\alpha$ to
\be
\frac{\Delta\alpha}{\alpha}=2.5_{-0.5}^{+2.5}\, {\rm ppm}\,, \label{bestfa}
\ee
and that of the baryon-to-photon ratio to
\be
\eta_{10}=6.24\pm0.06\,.\label{bestfn}
\ee
We note that both of these values correspond to the best fit and to the range of values within $\Delta\chi^2=1$ of it and that the latter is still marginally incompatible with the Planck value. Although, given the aforementioned assumptions, the difference is not statistically significant.

%%%%%%%%%%%%%%%%%%%%%%%%%%%%%%%%%%%%%%%%%%%%%%%%%%%%%%%%%%%%%%%%%%%%%%%%%%%%%%
\begin{table*}
\caption{Constraints on $\Delta\alpha/\alpha$, $N_\nu$, and $\eta_{10}$ as well as $R$ and $S$ for the clocks for the various choices of primordial abundances used in the analysis. The listed values correspond to the best fit in each case and to the range of values within $\Delta\chi^2=1$ of it (corresponding to the $68.3\%$ confidence level for a Gaussian posterior likelihood).}
\label{table8}
\centering
\begin{tabular}{c | c c c c c}
\hline
Abundances  & $\Delta\alpha/\alpha$ (ppm) & $N_\nu$ & $\eta_{10}$ & $R$ & $S$ \\
\hline
Baseline (No priors) & $5_{-1}^{+3}$ & $2.7\pm0.2$ & $6.6\pm0.2$ & $60_{-10}^{+50}$ & $170_{-50}^{+120}$\\
Baseline (With priors) & $2.5_{-0.5}^{+2.5}$ & $2.98\pm0.02$ & $6.24\pm0.06$ & $40_{-10}^{+30}$ & $100\pm30$\\
\hline
Null (No priors) & $0_{-2}^{+6}$ & $2.8\pm0.2$ & $6.0_{-0.2}^{+0.3}$ & $10_{-10}^{+90}$ & $20_{-20}^{+200}$\\
Null (With priors) & $0.8_{-0.6}^{+1.5}$ & $2.98\pm0.02$ & $6.08\pm0.06$ & $10_{-10}^{+20}$ & $20_{-20}^{+50}$\\
\hline
Extended (No priors) & $5_{-1}^{+3}$ & $2.7\pm0.2$ & $6.6\pm0.2$ & $60_{-10}^{+50}$ & $170_{-50}^{+120}$ \\
Extended (With priors) &  $2.5_{-0.5}^{+2.5}$ & $2.98\pm0.02$ & $6.24\pm0.06$ & $40_{-10}^{+30}$ & $100\pm30$\\
\hline
\end{tabular}
\end{table*}
%%%%%%%%%%%%%%%%%%%%%%%%%%%%%%%%%%%%%%%%%%%%%%%%%%%%%%%%%%%%%%%%%%%%%%%%%%%%%%

%%%%%%%%%%%%%%%%%%%%
%%%%%%%%%%%%%%%%%%%%

\begin{figure*}
\centering
\includegraphics[width=8cm]{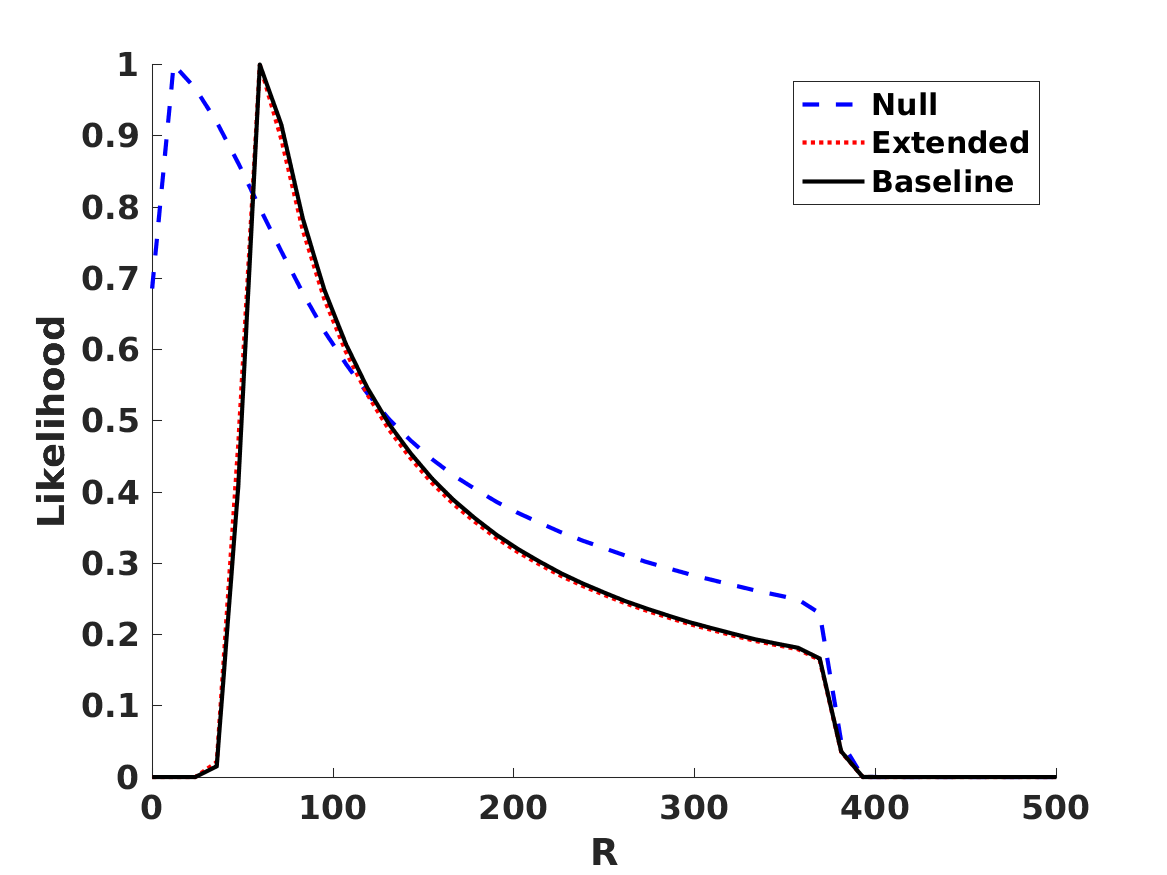}
\includegraphics[width=8cm]{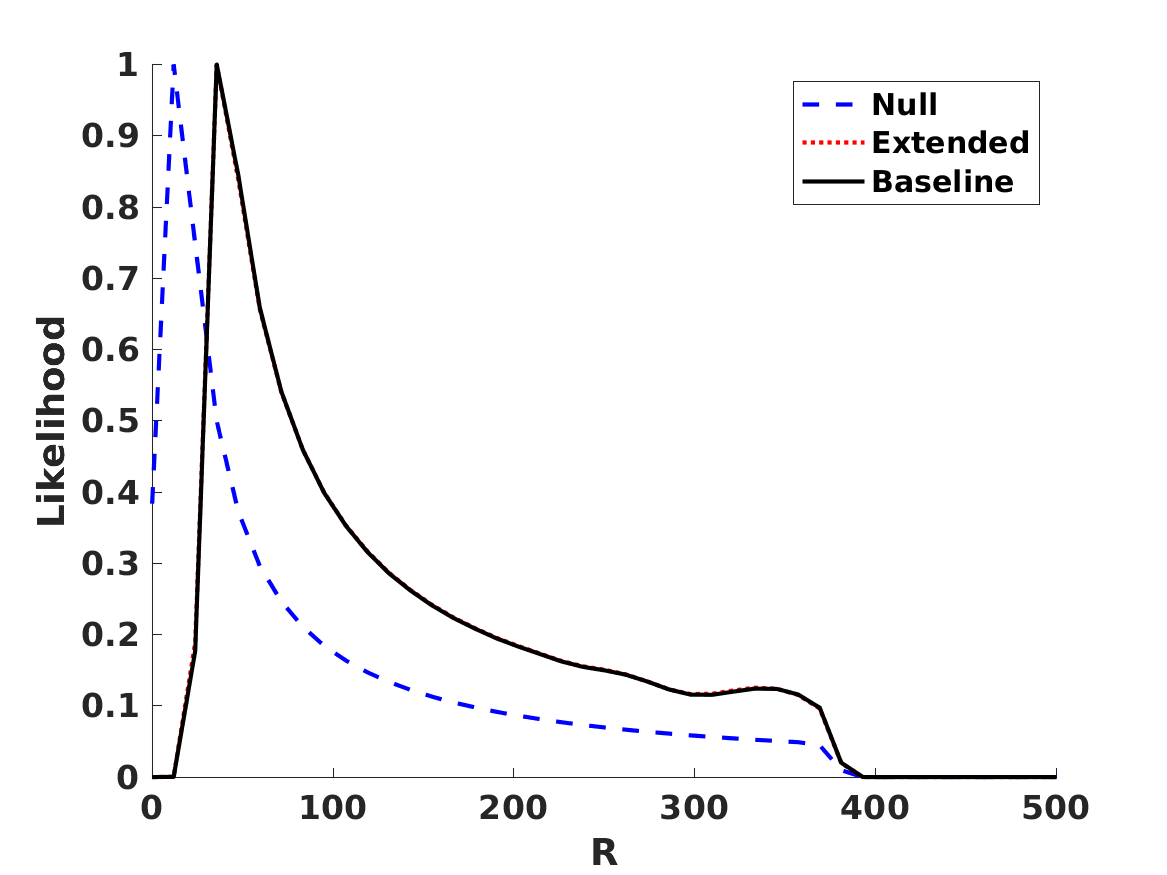}
\includegraphics[width=8cm]{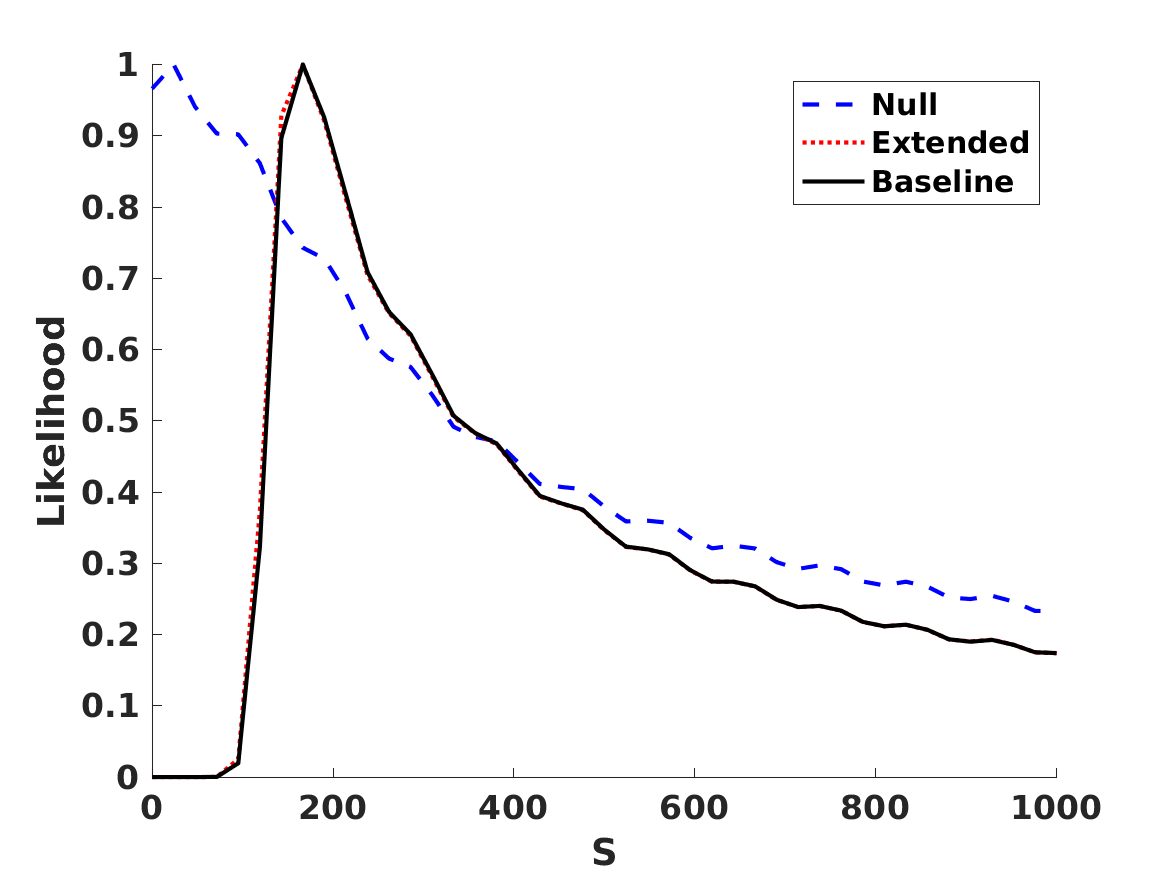}
\includegraphics[width=8cm]{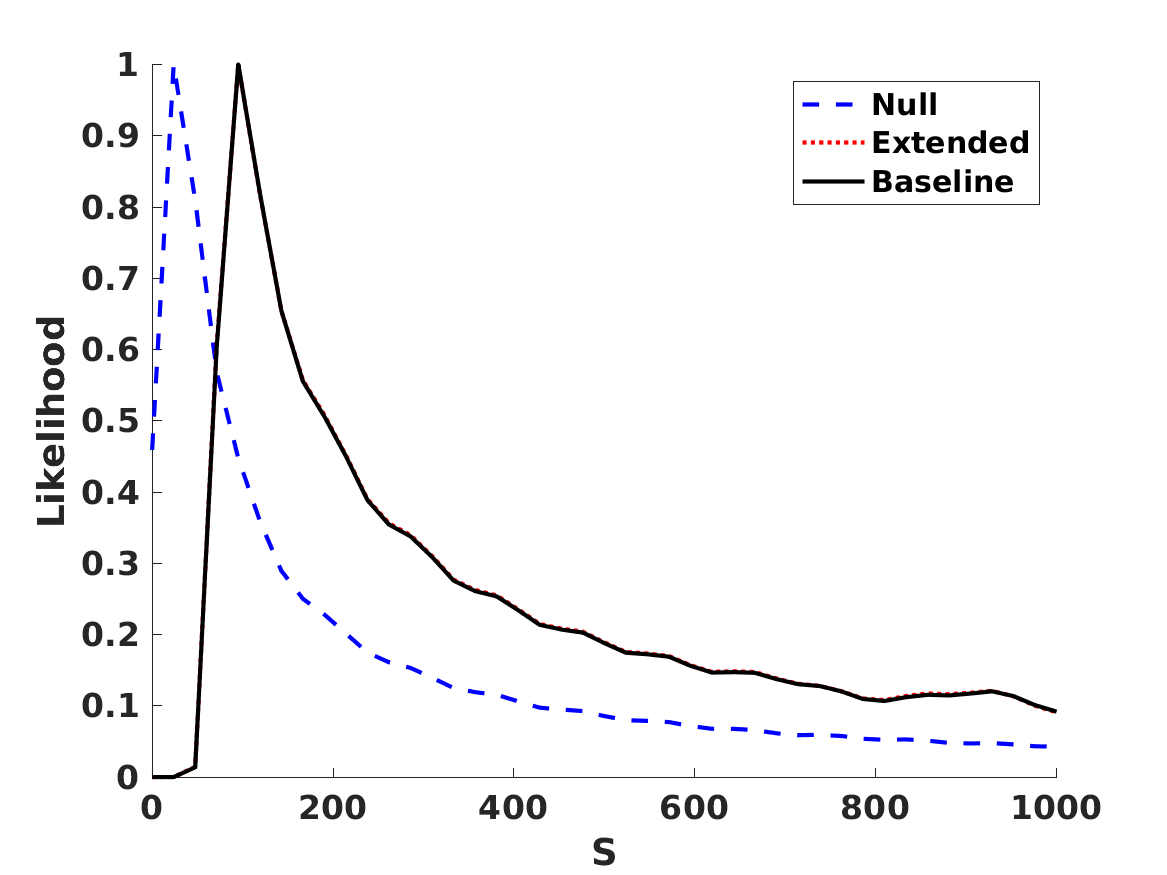}
\caption{One-dimensional posterior likelihoods for the parameters $R$ and $S$ (with the other parameters marginalised) for the clocks model. Left and right panels correspond to the cases without and with the two priors discussed in the text.}
\label{figure5}
\end{figure*}

Finally, in the case of the clocks model, one can also marginalise the three cosmological parameters and extract one-dimensional posterior likelihoods for the phenomenological parameters $R$ and $S$. While the two parameters were varied independently, the atomic clocks prior (cf. Eq.(\ref{clopri})) is a very stringent one, and the posterior likelihood in the $R$--$S$ parameter space effectively recovers this prior. For this reason, we do not show the constraints in this two-dimensional plane, but instead we show the corresponding one-dimensional likelihoods, which are plotted in Figure \ref{figure5}, while the derived constraints are also listed in  Table \ref{table8}. Given the various degeneracies and the difficulty in accurately sampling a five-dimensional parameter space, we only report constraints on these parameters to one significant digit.

Nevertheless, it is encouraging that the preferred values of both parameters, of
\begin{equation}
R\sim60\,,\quad S\sim170\,
\end{equation}
for the case without the LEP and Planck priors, or
\begin{equation}
R\sim40\,,\quad S\sim100\,
\end{equation}
for the case with the priors, are physically realistic. They are also, roughly speaking, in between those for the unification and dilaton models, which is qualitatively consistent with the previously discussed behaviour of the cosmological parameters in the clocks model by comparison to their behaviour in the other two models.

%%%%%%%%%%%%%%%%%%%%
%%%%%%%%%%%%%%%%%%%%

%%%%%%%%%%%%%%%%%%%%%%%%%%%%%%%%%%%%%%%%%%%%%%%%%%%%%%%%%%%%%%%
\section{Conclusions}
\label{concl}

We have extended the recent work of  \citet{Clara}, in which a self-consistent perturbative analysis of the effects of variations in nature's fundamental constants on BBN was carried out in the context of a broad class of GUT scenarios. The earlier work reports that a ppm level of relative variation in the fine-structure constant, $\alpha$, with a value at the BBN epoch that is larger than the present-day one is a possible solution of the lithium problem. In the present work, we have confirmed that this result is robust to extensions of the parameter space, whether these extensions are on the side of nuclear physics (specifically including the neutron lifetime as a free parameter) or on the side of cosmology (specifically including the number of neutrinos and the baryon-to-photon ratio as free parameters).

An interesting outcome of this extended analysis is that it enables a quantitative recasting of the lithium problem in terms of differences between the values of relevant parameters preferred by the observed primordial abundances and those measured by other direct or indirect methods. In the case of the neutron lifetime, the observed BBN abundances would favour a value that is always smaller than the one measured locally through the bottle or beam methods, even when allowing $\alpha$ as a free parameter. How much smaller this is is model-dependent, specifically depending on the values of the phenomenological parameters $R$ and $S$ since these affect the degeneracies between the neutron lifetime and $\alpha$.

In the cosmological context, the two relevant parameters are the number of neutrino species and the baryon-to-photon ratio. In this larger parameter space, the model-dependence is further manifested in the fact that, depending on the specific model, the lithium problem can be recast as a preference for a larger baryon-to-photon ratio with the number of neutrinos unchanged (in the case of the unification model) or as a preference for a smaller number of neutrinos with the baryon-to-photon ratio unchanged (in the case of the dilaton model). In all these cases, a larger value of $\alpha$ at the BBN epoch is preferred. We again emphasise that these results are driven by the observed lithium abundance; if this is excluded from the analysis, then within the statistical uncertainties, the standard values are recovered throughout.

The effects of the broader parameter space on the baryon-to-photon ratio $\eta_{10}$ are particularly interesting. The lithium-7 abundance increases with increasing $\eta_{10}$ in the relevant region of parameter space, so one might expect that increasing $\eta_{10}$ would worsen the lithium problem. On the other hand, the lithium-7 abundance decreases more strongly with increasing alpha. The overall result of the two effects is that a somewhat larger $\eta_{10}$ can be, so to speak, a statistically small price to pay to have a larger $\alpha$, overall decreasing the theoretically predicted abundance. This can be seen in the middle panels of Figures \ref{figure2}-\ref{figure4}. Clearly this is a model-dependent effect because the lithium-7 sensitivity to $\alpha$ depends on the parameters $R$ and $S$. Indeed the aforementioned effect on $\eta_{10}$ is most clearly seen in the case of the unification model (cf. Figure \ref{figure2}). It is therefore particularly instructive to contrast this with the case of the dilaton model (cf. Figure \ref{figure3}) where, in the absence of priors, a slightly smaller $\eta_{10}$ and a significantly smaller $N_\nu$ would be preferred. This highlights the potential of BBN as a tool to discriminate between different particle physics scenarios.

Overall, we conclude that the result that the observed BBN abundances prefer a ppm level variation of $\alpha$ is, despite the quantitative model dependence that we have highlighted, a qualitatively robust one. We also confirm, in agreement with CM20, that the helium-3 measurement (which is local rather than cosmological) has no significant impact in the analysis. We emphasise that ppm-level constraints at the BBN epoch are among the most stringent current constraints on possible $\alpha$ variations. For example, they are about three orders of magnitude stronger than current constraints from the comic microwave background, where one typically makes the simplifying assumption that only $\alpha$ varies, while the rest of the underlying physics is unchanged. That said, we expect that under the same GUT paradigm assumptions, it should also be possible to constrain $\alpha$ to the ppm level of sensitivity using the cosmic microwave background.

By excluding lithium from the analysis, which we dubbed the null case, we also obtained upper limits on possible variations of $\alpha$ at the BBN epoch. These can be thought of as BBN constraints on $\alpha$ on the assumption that the solution to the lithium problem is an astrophysical one. At the two-sigma level, our exploration of the parameter space suggests a bound $|\Delta\alpha/\alpha|<50$ ppm without any priors, but such comparatively large variations are normally associated with non-standard values of the neutron lifetime, the number of neutrinos, or the baryon-to-photon ratio. By including the local experimental constraints on the first two of these parameters and the cosmic microwave background constraint on the third parameter as priors, and further including the atomic clocks prior with the combination of the unification parameters $R$ and $S$, the constraint also tightens to $|\Delta\alpha/\alpha|<5$ ppm at the two-sigma level.

We also note that this ppm level sensitivity at the BBN epoch is comparable to the one obtained with low-redshift ($z\leq4.5$) UV-optical spectroscopic measurements: Sub-ppm constraints are extremely rare \citep{Kotus}, and even ppm ones are uncommon for single targets \citep{ROPP}. Recent advances in infrared spectroscopy have enabled measurements in the redshift range $5\leq z\leq7$, but only with a sensitivity of tens of ppm \citep{Wilczynska}. That said, these spectroscopic measurements are direct and model-independent and their sensitivity is expected to improve with the ESPRESSO spectrograph \citep{Leite}. In any case, this comparison does mean that such a putative ppm level variation at the BBN epoch is fully compatible with current low-redshift and local laboratory measurements.

Our analysis confirms BBN as a sensitive probe of new physics. While, as we have made explicit, there is an inherent model dependence in this BBN analysis, and there is a positive aspect to it. Should varying fundamental constants be confirmed by future observations or experiments, BBN would become a powerful probe of GUT scenarios. Importantly, this will require not only precise and accurate determinations of the primordial abundances, but also sensitive measurements of other quantities that affect BBN, including the neutron lifetime, number, or neutrinos, and baryon-to-photon ratio.

In conclusion, our analysis highlights a possible and physically motivated solution to the lithium problem. While one can legitimately argue that the simplest solution to the lithium problem is to be found within observational astrophysics, our work shows that varying fundamental constants provides a viable alternative.

%%%%%%%%%%%%%%%%%%%%%%%%%%%%%%%%%%%%%%%%%%%%%%%%%%%%%%%%%%%

\begin{acknowledgements}
This work was supported by FCT---Funda\c c\~ao para a Ci\^encia e a Tecnologia through national funds (PTDC/FIS-AST/28987/2017) and by FEDER---Fundo Europeu de Desenvolvimento Regional funds through the COMPETE 2020---Operacional Programme for Competitiveness and Internationalisation (POCI-01-0145-FEDER-028987). Useful correspondence with Paolo Molaro is gratefully acknowledged. 
\end{acknowledgements}

\bibliographystyle{aa} % style aa.bst
\bibliography{bbn} % your references Yourfile.bib
\end{document}